\documentclass[pra,twocolumn,10pt,amsmath, superscriptaddress,notitlepage,showpacs]{revtex4-1}
\usepackage{amsmath,amsfonts,amssymb,amstext,amscd,amsthm,dsfont,bbm,hyperref,natbib}
\usepackage{physics}
\usepackage{color}
\usepackage{soul,xcolor}
\usepackage{gensymb}
\hypersetup{
    colorlinks = true,
    citecolor=blue,
    linkcolor = red,
    anchorcolor = red,
    citecolor = blue,
    filecolor = red,
    pagecolor = red,
    urlcolor = red,
    }
\usepackage[normalem]{ulem}
\usepackage{graphicx}
\usepackage{bbold}
\usepackage{braket}
\usepackage{placeins}

\begin{document}
\title{An Invitation to Quantum Channels}
	\author{Vinayak Jagadish}
	\email{jagadishv@ukzn.ac.za}
	\affiliation{Quantum Research Group, School  of Chemistry and Physics,
		University of KwaZulu-Natal, Durban 4001, South Africa}\affiliation{ National
		Institute  for Theoretical  Physics  (NITheP), KwaZulu-Natal,  South
		Africa}
	\author{Francesco Petruccione}
	\email{petruccione@ukzn.ac.za}
	\affiliation{Quantum Research Group, School  of Chemistry and Physics,
		University of KwaZulu-Natal, Durban 4001, South Africa}\affiliation{ National
		Institute  for Theoretical  Physics  (NITheP), KwaZulu-Natal,  South
		Africa}
	
\begin{abstract} 
Open quantum systems have become an active area of research, owing to its potential applications in many different fields ranging from computation to biology. Here, we review the formalism of dynamical maps used to represent the time evolution of open quantum systems and discuss the various representations and properties of the same, with many examples.
\end{abstract}
\maketitle
\section{Introduction}
\label{introduction}
Quantum systems being in complete isolation are very uncommon. Generally an open quantum system\cite{breuer2007theory} is one which is interacting with its surroundings. In order to investigate the properties of such a quantum system, e.g., its dynamics, we need to know the effects of the environment on the system. The environment by definition is an external system about which we have little or no information. In this situation, one would therefore like to develop a description of open quantum systems described by a dynamical equation that accounts for the influence of the surrounding environment on the system state, while removing the need to track the full environment evolution. 
The study of evolution of open quantum systems is a problem of great excitement that has led to the development of different approaches, each having their merit based on the context of study. 
In this article, we consider the finite time evolution of open systems via the concept of dynamical maps, in detail. By finite time evolution, it is to understood as the change of the state of a system from an arbitrary initial instant of time to a later one.

Any quantum system whose state space dimension is finite may be represented by an $n \times n$ matrix, called the density matrix $\rho$. The density matrix should be of unit trace, Hermitian and positive.
\begin{equation}
{\mbox{tr}}(\rho) = 1 \quad ; \quad \rho^{\dagger} = \rho \quad ; \quad \langle x | \rho |x \rangle \geq 0 \thinspace \forall x .\label{constraints}
\end{equation}

For closed quantum systems, the purity of a quantum state is not affected under time evolution and the evolution is unitary. However, for open quantum systems, the purity of quantum states can decrease in principle and one has to account for mixed states as well. Hence for open quantum systems, it becomes necessary to represent states using density matrices.

The evolution of an open system may be stochastic opposed to being deterministic for closed systems. Open quantum systems usually have temporal evolutions that are not unitary and cannot be described by a Hamiltonian scheme. In the following sections, we discuss the formalism of the method of stochastic maps describing the finite time evolution of open quantum systems. 

We review the formalism of dynamical maps used to represent the time evolution of open quantum systems. This article is for beginners in the area of open quantum systems who find the different representations and their transformations confusing. We motivate the ideas from a historic perspective, but in a pedagogical manner.

The plan of the article is as follows. We start with closed evolution of the total system, i.e., the system and environment taken together and obtain the finite time evolution of the state of the system by tracing out the environment. Then, we introduce finite-time maps in an abstract way, outline the various representations of the same and then show how to connect the various representations. We exemplify the maps on a qubit in some detail and then look briefly into maps which are not completely positive. Finally, we present the differential form of maps as well.

\section{Time evolution of open quantum systems}
\label{timeevolu}
Let us start with closed evolution, which we are familiar with. We have a system of interest whose states live in the Hilbert space $\mathcal{H}_S$. Now, we have the environmental Hilbert space, $\mathcal{H}_E$ as well. We extend the Hilbert space as $\mathcal{H} = \mathcal{H}_S \otimes \mathcal{H}_E$ such that the total system is closed and the total evolution is unitary. If the initial states of the system and environment are $\rho_{S}(0)$ and $\rho_{E}(0)$, respectively, and they are uncorrelated initially, the evolution under the total unitary operator $U(t)$ is 
\begin{equation}
	\rho^\prime (t) = U(t)\rho_{S}(0)\otimes\rho_{E}(0) U^{\dagger}(t), \,\!  \label{unitary1}
\end{equation}
where $\rho^\prime (t)$ denotes the state of the total system at a time~$t$. The assumption that the initial state of the composite system is separable looks restrictive, but since one can have good control on the system, this assumption is not a bad one. Also, since $t = 0$ is assumed to be the time at which the system and the environment start interacting, it makes sense to assume that they are not correlated before interacting. Now, to obtain the state of the system at time~$t$, we do a partial trace over the environment. 
	\begin{equation}
	\rho_{S} (t) = \text{tr}_{E} \Big(U(t) [\rho_{S}(0)\otimes\rho_{E}(0)] U^{\dagger}(t)\Big). \,\!  \label{envrep1}
	\end{equation}
	Let the dimension of the system be $n$ and that of the environment be $m$ such that $m\gg n$.  Choosing a basis $\{|k\rangle\}$ for the environment and taking the initial state of the environment as mixed, $\rho_{E}(0) = \sum_{k} p_{k} |k\rangle\langle k|$ and noting that the basis is orthonormal, i.e. $\langle k| k'\rangle = \delta_{k k'}$, we obtain
	\begin{eqnarray}
	\label{parttrace1}
	\rho_{S} (t) &=& \sum _k \langle k | U(t)\thinspace \rho_{S}(0)\otimes\rho_{E}(0)\thinspace U^{\dagger}(t) | k \rangle \nonumber\\
	&=&  \sum _{k,k'}  p_{k' }\langle k | U(t) | k' \rangle  \rho_{S}(0)\langle k'| U^{\dagger}(t) | k \rangle. 
	\end{eqnarray}
	One can see that $U(t)$ is an $nm \times nm$ matrix and $\langle k | U(t) | k' \rangle$ is therefore an $n \times n$ matrix. It is not the usual matrix element, which being a number, rather it is a matrix and it is often called the \emph{reduced matrix element}. 
Denoting $\sqrt{p_{k'}} \langle k | U(t) | k' \rangle$ as $D_{k,k'}$ , we can write Eq.~(\ref{parttrace1}) as
\begin{eqnarray}
 \label{krausrep2}
	\rho_{S} (t) &=& \sum_{k,k'} D_{k,k'}^{}\rho_{S}(0) D_{k,k'}^\dagger \nonumber\\
	&=& \sum_\alpha D_{\alpha}\rho_{S}(0) D_{\alpha}^\dagger.
	\end{eqnarray}
	where we have combined the two indices $kk'$ into one index, $\alpha$. The condition that $\rho_{S} (t)$ is a valid density matrix with trace 1 sets the following condition.
	\begin{equation}
 \sum_{\alpha} D_\alpha^\dagger D_\alpha^{\vphantom{\dagger}} = \mathbb{1}.
	\label{eqn:krausform1}
		\end{equation}
This is the \emph{trace-preserving} condition for the operators~$ D_\alpha$ and follows straight from the cyclic property of the trace.

\section{Quantum channels}
\label{quantummaps}

In the previous section, we have seen that the evolution of the system and environment taken together is given by a total unitary evolution operator $U$. Thus, for a closed dynamics, the discrete time evolution can be thought of as a unitary map $U: \rho(0) \to \rho^\prime(t)$.  In an analogous way, one can describe the open evolution by a map $\mathcal{E}$. A quantum channel (or map) $\mathcal{E}(\rho)$ on the system maps the initial quantum state of the system to the final state after its interaction with the environment. The map $\mathcal{E}(\rho)$ describes the finite time open dynamics of a quantum system. Any dynamical operation $\mathcal{E}$ on states of a quantum system $S$ must have the following properties:

(a) $\mathcal{E}$ must be linear. Hence it should respect the superposition principle of quantum mechanics.

(b) $\mathcal{E}$ should be Hermiticity preserving, which means that observables should be transformed into bonafide observables.

(c) $\mathcal{E}$ should be positivity as well as trace-preserving such that each density matrix should be transformed into a bonafide density matrix. 

If these three conditions are met, the map is said to be  positive and trace-preserving. A quantum system can in principle be a collection of many systems. Hence one needs to stipulate that if the map acts only on any one of the subsystem of the entire system, positivity should be preserved. On expanding the state of the system to a bigger composite one, the map is $\mathcal{E} \otimes \mathbb{1}^{(d)}$ (where $d$ is the dimension of the auxillary system) and the action of this map should also produce a valid state. $\mathcal{E}$ is called $d$-positive, if it is positive for the extension of the form $\mathcal{E}\otimes \mathbb{1}^{(d)}$. If the map remains positive for any $d$, then the map is said to be \emph{completely positive}. Therefore, on top of the three conditions mentioned above, one has to add the condition of \emph{complete positivity} (CP) for a valid map. It should be remembered that positivity of the map is a statement about its action on density matrices while complete positivity can be regarded as a statement about the map itself.

The map $\mathcal{E}$ can be represented in many ways, each of which have their own use depending on the context. We shall review the various representations in the subsequent sections. 

Before that, let us introduce the notations with a slight taste of mathematical rigor. $\mathcal{E}: \rho (t_0) \to \rho'(t_f)$, $\mathcal{E}$ denotes the abstract map (or quantum channel) that takes the state of the system, $\rho (t_0)$ at any arbitrary instant of time to that at a later time, $\rho'(t_f)$. In other words, the map $\mathcal{E}$ takes a state at any time $t_0$ and gives back the state at a later time $t_f = t_0 + t$, where $t$ is fixed (if one changes $t$, it changes the map). In the discussion that follows, $\rho'$ stands for the output state of a map, unless otherwise mentioned. Technically,  $\mathcal{E}: \mathcal{B}(\mathcal{H}_{S}) \to\mathcal{B}(\mathcal{H}_{S'})$ where $\mathcal{B}(\mathcal{H}_{S})$ and $\mathcal{B}(\mathcal{H}_{S'})$ denote the bounded operators of the input and output Hilbert spaces of finite dimensions (but can be different) respectively. In the context of maps acting on quantum states of the system, let $\mathcal{B}(\mathcal{H}_{S})$ and $\mathcal{B}(\mathcal{H}_{S'})$ denote the bounded operators of the input and output spaces of the system $S$. The map $\mathcal{E}: \mathcal{B}(\mathcal{H}_{S}) \to\mathcal{B}(\mathcal{H}_{S'})$ is completely positive if $\mathcal{E} \otimes \text{id}_{A}$ remains positive for all possible extensions. $\text{id}_{A}$ denotes the identity operator acting on operators $\mathcal{B}(\mathcal{H}_{A})$, the bounded operators of the auxillary system of Hilbert space $\mathcal{H}_{A}$.

\subsection{$\mathcal{A}-$form and $\mathfrak{B}-$form}

Sudarshan, Mathews and Rau \cite{sudarshan61a} considered the above mentioned three conditions as the defining conditions for the most general
quantum dynamical operation,
\begin{equation}
\rho'(t) = \mathcal{A}(t) \rho(0). 
\end{equation}
The linearity of $\mathcal{A}( t)$ stems from the linearity of quantum mechanics. The operator $\mathcal{A}$ can be written as a matrix if $\rho$ is finite dimensional and the transformation can be written as
\begin{equation}
 \rho_{ij} \longrightarrow \mathcal{A}_{ij;i^{\prime}j^{\prime}} \rho_{i^{\prime}j^{\prime}} = (\mathcal{A}\rho)_{ij} .
\end{equation}
In the equation given above, the elements of the density matrix (say $N$ dimensional) have been rearranged into a column vector so that $\mathcal{A}$ is an $N^2 \times N^2$ matrix. The index notation might look intimidating, but we use it here as a token of respect to the celebrated 1961 paper \cite{sudarshan61a}, which unfortunately did not receive proper attention, but which had in it almost every detail on finite time maps, although in disguise. We shall make the notation clear with an example for a map on a qubit and later on, we skip the index notation and show the necessary mathematical details as to how they come about.  
A general linear map relates the elements $\rho_{ij}$ of the input state to that of the output state $\rho'_{i'j'}$ as
\begin{equation}
\label{defA}
\rho'_{i'j'} = \sum_{i,j =1}^{N}\, \mathcal{A}_{i'j';ij}(t)\, \rho_{ij}, \ \ \  i',j'=1,2,\ldots, N.
\end{equation} 
The $\mathcal{A}$ map should ensure the preservation of hermiticity, i.e., $\rho^{\prime}_{i^{\prime}j^{\prime}} = \rho^{\prime *}_{j^{\prime}i^{\prime}}$,  and that of trace, i.e., $\text{ tr}[\rho^{\prime}]=1$. These imply that $\mathcal{A}$ have the following properties.
\begin{eqnarray}
\label{arest}
\mathcal{A}_{j^{\prime}i^{\prime};ji}=&\mathcal{A}^{*}_{i^{\prime}j^{\prime};ij} &{\mbox{(Hermiticity)}},\nonumber \\
\sum_{i^{\prime}}\,\mathcal{A}_{ i^{\prime}i^{\prime};ij}=&\delta_{i,j} &{\mbox{(Trace)}}.
\end{eqnarray}  

They also introduced a new matrix, $\mathfrak{B}$\,\!, which is related to $\mathcal{A}$\,\! by reshuffling, $\mathfrak{B}$  such that
\begin{equation}
  \label{Eq.rearrange}
 \mathfrak{B}_{ i^{\prime} i ; j^{\prime}j} = \mathcal{A}_{i^{\prime} j^{\prime};ij}.  
\end{equation}
In the form $\mathfrak{B}$ the restrictions in Eq.~(\ref{arest}) on the map gets modified to the following relations:
\begin{eqnarray}
\label{brest}
\mathfrak{B}_{j^{\prime}j;i^{\prime}i} =& \mathfrak{B}^*_{i^{\prime}i;j^{\prime}j}  &{\mbox{(Hermiticity)}},\nonumber \\
\sum_{i^{\prime}}\, \mathfrak{B}_{i^{\prime}i; i^{\prime}j} \; =& \delta_{ij}  \quad \quad &{\mbox{(Trace)}} 
\end{eqnarray}
The operator $\mathfrak{B}$ represents the most general transformation that a quantum system can undergo and so is referred to as the {\em dynamical map} or {\em dynamical matrix}. One has to note that $\mathfrak{B}$ is a Hermitian matrix whereas $\mathcal{A}$ is not. It is not necessary that $\mathfrak{B}$ itself be a positive matrix for maintaining the positivity of the density matrices under dynamical evolution. If $\mathfrak{B}$ is in itself a positive matrix, the map is completely positive, which we will show later. 

Let us try and understand the $\mathcal{A}$ and $\mathfrak{B}$ representations of the map for the case of a qubit. Let \begin{equation}
  \rho =  \left( \begin{array}{cc}\rho_{00} & \rho_{01}\\ \rho_{10} & \rho_{11} \end{array} \right) \end{equation} 
  denote the density matrix of a qubit in the standard (computational) basis, $\{|0\rangle, |1\rangle\}$, $|0\rangle = (1, 0)^{T}$ and $|1\rangle = (0, 1)^{T}$. Then $\rho' = \mathcal{A} \rho$ can be understood as follows.
 \begin{equation}
\begin{pmatrix}
  \rho'_{00} \\
  \rho'_{01} \\
  \rho'_{10} \\
  \rho'_{11} \end{pmatrix} = \begin{pmatrix}
 \textcolor {red}{\mathcal{A}_{00;00}} & \textcolor {red}{\mathcal{A}_{00;01}} & \textcolor {red}{\mathcal{A}_{00;10}} & \textcolor {red}{\mathcal{A}_{00;11} }\\
  \textcolor {blue}{\mathcal{A}_{01;00}} & \textcolor {blue}{\mathcal{A}_{01;01}} & \textcolor {blue}{\mathcal{A}_{01;10}} & \textcolor {blue}{\mathcal{A}_{01;11} } \\
  \textcolor {green}{\mathcal{A}_{10;00}} & \textcolor {green}{\mathcal{A}_{10;01}} & \textcolor {green}{\mathcal{A}_{10;10}} & \textcolor {green}{\mathcal{A}_{10;11} }\\
   \textcolor {black}{\mathcal{A}_{11;00}} & \textcolor {black}{\mathcal{A}_{11;01}} & \textcolor {black}{\mathcal{A}_{11;10}} & \textcolor {black}{\mathcal{A}_{11;11} }
\end{pmatrix} \begin{pmatrix}
  \rho_{00} \\
  \rho_{01} \\
  \rho_{10} \\
  \rho_{11} \end{pmatrix}.
\end{equation} The $\mathfrak{B}$ matrix obtained by reshuffling $\mathcal{A}$ is hence
 \begin{equation}
\mathfrak{B} = \begin{pmatrix}
 \textcolor {red}{\mathcal{A}_{00;00}} & \textcolor {red}{\mathcal{A}_{00;01}} &  \textcolor {blue}{\mathcal{A}_{01;00}} & \textcolor {blue}{\mathcal{A}_{01;01}}\\
  \textcolor {red}{\mathcal{A}_{00;10}} & \textcolor {red}{\mathcal{A}_{00;11} } & \textcolor {blue}{\mathcal{A}_{01;10}} & \textcolor {blue}{\mathcal{A}_{01;11} } \\
  \textcolor {green}{\mathcal{A}_{10;00}} & \textcolor {green}{\mathcal{A}_{10;01}} &  \textcolor {black}{\mathcal{A}_{11;00}} & \textcolor {black}{\mathcal{A}_{11;01}} \\
   \textcolor {green}{\mathcal{A}_{10;10}} & \textcolor {green}{\mathcal{A}_{10;11} }& \textcolor {black}{\mathcal{A}_{11;10}} & \textcolor {black}{\mathcal{A}_{11;11} }
\end{pmatrix}. 
\end{equation}
The elements in the four rows of $\mathcal{A}$ are presented in different colors to bring clarity as to how the rows are being \emph{folded} to write the dynamical matrix $\mathfrak{B}$. 

\subsection{Operator-sum representation and the process matrix}

We shall look into another representation of the map, stemming from $\mathfrak{B}$ here.
$\mathfrak{B}$\,\! is a Hermitian matrix and thus admits a spectral decomposition. Let $\lambda_\alpha$\,\! be the eigenvalues and the corresponding eigenvectors be $|\Lambda^{(\alpha)}\rangle$\,\!. Then \begin{equation}
  \label{Eq.Bdecomp1}
  \rho'_{i'j'}= \sum_{\alpha} \lambda_{\alpha} \Lambda_{i'i}^{(\alpha)} \rho_{ij} \Lambda_{j'j}^{(\alpha)\dagger},
\end{equation}
For brevity, let us define $\Lambda^{(\alpha)}_{i'i} = C_{\alpha}$\,\!. Then, 
\begin{equation}
\rho^\prime = \sum_\alpha \eta_\alpha C_{\alpha}^{}\rho C_{\alpha}^\dagger. \,\!
\end{equation}
If all the eigenvalues $\lambda_\alpha$\,\! of $\mathfrak{B}$ are positive, then one can define $D_{\alpha}= \sqrt{\lambda_\alpha}\; C_{\alpha}$\,\!, which allows the map to be written as
\begin{equation}
	\rho^\prime = \sum_\alpha D_{\alpha}^{}\rho D_{\alpha}^\dagger. \,\!  \label{krausrep}
	\end{equation}
This is called the Operator-Sum Representation (OSR) or Kraus decomposition and each of the $D_{\alpha}$'s are called Kraus operators. This was independently discovered by Kraus in \cite{kraus_general_1971}, 10 years later, but was already discussed in the celebrated 1961 paper \cite{sudarshan61a}. If the map admits an OSR, then it is equivalent to the fact that the $\mathfrak{B}$~matrix is positive. The map in the Kraus (operator-sum) form is trace preserving if
\begin{equation}
 \sum_{\alpha} D_\alpha^\dagger D_\alpha^{\vphantom{\dagger}} = \mathbb{1}.
	\label{eqn:krausform}
\end{equation}

The operator sum decomposition is not unique. There is a unitary freedom in choosing the operators in Eq.~(\ref{krausrep}). 
Two different sets of Kraus operators $\{ D_\alpha \}$\,\! and $\{K_\beta \}$\,\! related by a unitary transformation reproduce the same open evolution of $\rho_S$\,\!.
Each of the Kraus operators for a map~$\mathcal{E}$ can be expanded in a suitable operator basis $\{A_{i}\}$ as $D_{\alpha} = \sum_i a^{(\alpha)}_{i} A_{i},$ with $a^{(\alpha)}_{i} \in \mathbb{C}$. The operator basis can be chosen to be orthonormal (tr$[ A_{j}^{\dagger} A_k^{\vphantom{\dagger}} ] = \delta_{jk}$) for convenience. 
\begin{equation}
	\mathcal{E}(\rho) = 
	 \sum_{ij}\chi_{ij} A_i \rho A_j^\dagger, \quad\text{where}\quad \chi_{ij}=\sum_{\alpha} a_i^{(\alpha)}a_j^{(\alpha)*}.
    \label{eqn:chikraus}
\end{equation}
So for a given basis set $\{A_i \}$ the matrix $\chi$ completely characterizes $\mathcal{E}$.  The $\chi$ matrix is Hermitian and different Kraus representations of the same process $\mathcal{E}$ have the same $\chi$ matrix. The $\chi$ matrix is also called the process matrix in the context of quantum process tomography.

\subsection{Complete positivity and operator-sum representation}

Having spoken about complete positivity, let us now see how the representation of the map reveals it.
We can easily show that the map represented in the operator-sum form implies complete positivity. Let us recall that for complete positivity, $\mathcal{E} \otimes \text{id}_{A}$ is positive. If $|\phi\rangle$ denotes the state of the composite system (system and the auxillary taken together) and $P$ is any positive operator acting on it, CP implies the following inequality
\begin{equation}
	\langle\phi|(\mathcal{E} \otimes \text{id}_{A}) P |\phi\rangle \geq 0.
\end{equation}
Invoking the OSR for $\mathcal{E}$, 
\begin{eqnarray}
\langle\phi|(\mathcal{E} \otimes \text{id}_{A}) P |\phi\rangle 
&=&\sum_{\alpha} \langle\phi|(D_{\alpha}\otimes \mathbb{1}) P (D_{\alpha}^\dagger\otimes \mathbb{1})  |\phi\rangle \nonumber\\ 
&=&\sum_{\alpha} \langle \phi_{\alpha}|P|\phi_{\alpha}\rangle,
\label{eqn:krauscpcondn}
\end{eqnarray}
where $|\phi_{\alpha}\rangle  = (D_{\alpha}^\dagger\otimes \mathbb{1})  |\phi\rangle$. Since $P$ is positive, the RHS of Eq.~(\ref{eqn:krauscpcondn}) is positive and hence it is verified that the OSR implies CP. $ \mathbb{1}$ is the identity matrix corresponding to $\text{id}_{A}$ and of same dimensions as of $D_{\alpha}$.

\section{Switching between representations and various properties}
\label{switching}

\subsection{Preliminaries}

Let us take a short mathematical digression, introducing a few notations and a few results \cite{dariano_bell_2000} which will help us to follow the subsequent ideas presented.
Let $X, Y, Z$ be operators from $\mathcal{H}_{S} \to\mathcal{H}_{S'}$. Let us assume that all are $n$~dimensional, for simplicity. The square matrices can be written as a vector, by stacking the elements row-by-row. The vectorized form is notated as $|.\rangle\rangle$. $|Z\rangle\rangle$ denotes the vectorized version of the operator $Z$ and is an element of $\mathcal{H}_{S}\otimes \mathcal{H}_{S}$. The following results are useful for our future discussion and are given without proof.  
\begin{eqnarray}
X\otimes Y |Z\rangle\rangle &=& |XZY^{T}\rangle\rangle, \label{identity1} \\
|Z\rangle\rangle &=& Z\otimes \mathbb{1} |I\rangle\rangle \langle\langle I|, \label{identity2} \\
\text{tr}_{\mathcal{H}_{S}} (|X\rangle\rangle \langle\langle Y|) &=& XY^{\dagger}, \label{identity3} \\
\text{tr}_{\mathcal{H}_{S'}} (|X\rangle\rangle \langle\langle Y|) &=& X^{T}Y^{*}. \label{identity4}
\end{eqnarray}
One should note that all these relations are basis-dependent and hence one must fix a basis throughout for convenience.

If 
\begin{equation}
Z = \left(\begin{matrix} 
	Z_{11} &  \dots&\dots Z_{1,n}\\
	\vdots    & \vdots    & \vdots \\
	Z_{n,1}  & \dots&\dots Z_{n,n}
\end{matrix}\right),
\end{equation} is an $n \times n$ matrix, vectorization stacks row by row to make a column, i.e. an $n^2 \times 1$ column matrix:
\begin{equation}
|Z\rangle\rangle =  \begin{pmatrix}
Z_{11} \\
 \vdots \\
 Z_{1,n}\\
 \vdots\\
  Z_{n,1} \\
  \vdots\\
  Z_{n,n}
\end{pmatrix}.
\end{equation}
The matrix $Z$ need not be square for it to be vectorized. Any matrix which is not columnar can be vectorized, but we use only square matrices for our discussion throughout.

One can also talk about undoing the vectorization. This is matricizing or the $\mathrm{mat}$ operation which stacks the elements of a column matrix row by row with rows of length $n$ to generate an $n \times n$ square matrix.  
\begin{equation}
	\mathrm{mat} |Z\rangle\rangle = Z.
	\end{equation}
Let $\{|e_i\rangle\}$ denote the standard basis of vectors in $n$ dimensional space. ($\{|e_i\rangle\}$ is a column of zeros with 1 at the $i$ th row.) Then, $I = \sum_{i}|e_i\rangle \langle e_i|$, 
\begin{eqnarray}
	 |I\rangle\rangle &=& \sum_{i}|e_i\rangle\otimes |e_i\rangle, \nonumber\\
	 |I\rangle\rangle \langle\langle I|&=& \left(\begin{matrix} 
	|e_1\rangle\langle e_1| &  \dots&\dots |e_1\rangle\langle e_n|\\
	\vdots    & \vdots    & \vdots \\
	|e_n \rangle\langle e_1|    & \dots&\dots|e_n \rangle\langle e_n| 
\end{matrix}\right).
\end{eqnarray}
Note that $ |I\rangle\rangle \langle\langle I|$, in 2 dimensions is nothing but $2 |\phi^{+}\rangle \langle \phi^{+}|$ where $|\phi^{+}\rangle = \frac{1}{\sqrt{2}}( |00\rangle + |11\rangle)$, one of the Bell-states, which is a maximally entangled state.

\subsection{Connecting $\mathcal{A}$ and Kraus form}

We start with the Kraus form of the map,
\begin{equation*}
	\rho^\prime = \sum_\alpha D_{\alpha}^{}\rho D_{\alpha}^\dagger.
	\end{equation*}
	Using Eq.~(\ref{identity1}), writing the vectorized version as
	\begin{eqnarray}
|\rho^\prime\rangle\rangle &=& \sum_\alpha |D_{\alpha}^{}\rho D_{\alpha}^\dagger\rangle\rangle \nonumber\\
&=& \Big(\sum_\alpha D_{\alpha} \otimes  D_{\alpha}^{*}\Big) |\rho \rangle\rangle. 
\end{eqnarray}
We know that $|\rho^\prime\rangle\rangle = \mathcal{A} |\rho\rangle\rangle$. Hence it follows that 
\begin{equation}
\mathcal{A} = \sum_\alpha D_{\alpha}\otimes D_{\alpha}^{*}.
\label{Ainkraus}
\end{equation}

\subsection{Connecting $\mathfrak{B}$ and Kraus form}

We now work in 2 dimensions for a while, for simplicity and showing explicit details. We therefore restrict ourselves to the case of a single term in Eq.~(\ref{Ainkraus}), i.e.,~$\mathcal{A} =  D_{1}\otimes D_{1}^{*}$, since the summation does not change the structure of the matrices. 

Let
 \begin{equation}
 D_1=  \left( \begin{array}{cc}D_{00} & D_{01}\\ D_{10} & D_{11} \end{array} \right).
 \end{equation} 
The $\mathcal{A}$ matrix is therefore
\begin{equation}
\mathcal{A} = \begin{pmatrix}
 \textcolor {red}{D_{00}D_{00}^*} & \textcolor {red}{D_{00}D_{01}^*} & \textcolor {red}{D_{01}D_{00}^*} & \textcolor {red}{D_{01}D_{01}^* }\\
  \textcolor {blue}{ D_{00}D_{10}^* }&  \textcolor {blue}{D_{00}D_{11}^* }&  \textcolor {blue}{D_{01}D_{10}^*} &  \textcolor {blue}{D_{01}D_{11}^*} \\
  \textcolor {green}{ D_{10}D_{00}^*} &  \textcolor {green}{D_{10}D_{01}^*} &  \textcolor {green}{D_{11}D_{00}^*} &  \textcolor {green}{D_{11}D_{01}^* }\\
   \textcolor {black}{D_{10}D_{10}^* }&  \textcolor {black}{D_{10}D_{11}^* }&  \textcolor {black}{D_{11}D_{10}^*} &  \textcolor {black}{D_{11}D_{11}^*}
\end{pmatrix}.
\label{atensprod}
\end{equation} 
By reshuffling, one can write the corresponding $\mathfrak{B}$ matrix as
\begin{equation}
\mathfrak{B} = \begin{pmatrix}
  \textcolor {red}{D_{00}D_{00}^* }& \textcolor {red}{D_{00}D_{01}^* }&  \textcolor {blue}{D_{00}D_{10}^*} &  \textcolor {blue}{D_{00}D_{11}^*} \\
  \textcolor {red}{D_{01}D_{00}^* }& \textcolor {red}{D_{01}D_{01}^*} & \textcolor {blue}{ D_{01}D_{10}^*} &  \textcolor {blue}{D_{01}D_{11}^*} \\
   \textcolor {green}{D_{10}D_{00}^*} &  \textcolor {green}{D_{10}D_{01}^* }&  \textcolor {black}{D_{10}D_{10}^*} & \textcolor {black}{ D_{10}D_{11}^*} \\
   \textcolor {green}{D_{11}D_{00}^*} &  \textcolor {green}{D_{11}D_{01}^*} &  \textcolor {black}{D_{11}D_{10}^*} &  \textcolor {black}{D_{11}D_{11}^*}
\end{pmatrix}.
\label{btensprod}
\end{equation}
Since $|D_{1}\rangle\rangle$, the vectorized version of the square matrix $D_1$ is
 \begin{equation*}
|D_{1}\rangle\rangle = \begin{pmatrix}
  D_{00} \\
  D_{01} \\
  D_{10} \\
  D_{11}
\end{pmatrix},
\end{equation*} it is clear that $\mathfrak{B} = |D_{1}\rangle\rangle  \langle\langle D_{1}|$. This means that the $\mathfrak{B}$~matrix is nothing but the sum of outer-products of the vectorized version of the Kraus matrices and therefore a Hermitian matrix, 
\begin{equation}
\mathfrak{B} = \sum_\alpha |D_{\alpha}\rangle\rangle  \langle\langle D_{\alpha}|.
\label{Binkraus}
\end{equation}
Now, one can easily deduce that the Kraus operators are the matricized versions of the eigenvectors of $\mathfrak{B}$. The $\mathfrak{B}$~matrix being Hermitian admits a spectral decomposition.  Letting $\lambda_\alpha$\,\! and $|\Lambda^{(\alpha)}\rangle$\,\! to be the corresponding eigenvalues and eigenvectors, respectively,
 \begin{equation}
  \label{Eq.Bdecomp2}
 \mathfrak{B} = \sum_{\alpha} \lambda_{\alpha} |\Lambda^{(\alpha)}\rangle \langle \Lambda^{(\alpha)}| ,
\end{equation}
it is apparent that $D_{\alpha} = \sqrt{\lambda_{\alpha}}\thinspace \mathrm{mat}  |\Lambda^{(\alpha)}\rangle$, where $\mathrm{mat}$ refers to matricizing as explained above. {The extension to arbitrary dimensions is thus straightforward. For an $N\times N$~dynamical matrix, the eigenvectors are column matrices with $N$ rows. The Kraus operators are written down by stacking the elements of the eigenvector matrix (column matrix) row by row with rows of length $N$ and multiplied by the square-root of the corresponding eigenvalue.}

It should be carefully noted that positivity of $\mathfrak{B}$ means that all of its eigenvalues $\lambda_{\alpha}$ are positive. As we shall see later, $\mathfrak{B}$ can have negative eigenvalues which leads to an operator-sum-difference representation.

\subsection{Connecting $\mathcal{A}$ and $\mathfrak{B}$ via the Kraus form}

\label{abindexkraus}
We have seen the different representations of the map. Now, we write the details with indices as in \cite{sudarshan61a} so that all the ideas are connected together. Let us recall Eq.~(\ref{krausrep}).
\begin{equation*}
	\rho^\prime = \sum_\alpha D_{\alpha}^{}\rho D_{\alpha}^\dagger. 
	\end{equation*}
		Representing the same in terms of indices (remember that all indices run from $1...N$ for an $N$ dimensional density matrix), this is
	\begin{equation}
\label{Akrausindex}
\rho'_{i'j'} = \sum_\alpha (D_{\alpha})_{i'i} (D_{\alpha})_{j'j}^{*}\thinspace  \rho_{ij}.
\end{equation} 
Remember that the repeated indices are summed over. Using Eq.~(\ref{Ainkraus}), 
\begin{equation}
\sum_\alpha (D_\alpha)_{i'i} (D_\alpha)_{j'j}^{*} = (\sum_\alpha D_\alpha\otimes D_\alpha^{*})_{i'j';ij},
\end{equation} we can connect it with the definition of $\mathcal{A}$ as in Eq.~(\ref{defA}). 
The Eq.~(\ref{Binkraus}) tells us that $\mathfrak{B}$ is the outer-product of $D$~matrix. We therefore write the $\mathcal{A}$ and $\mathfrak{B}$ forms in terms of the matrix elements of the Kraus matrix, labelled by indices as
\begin{eqnarray}
\mathcal{A}_{i'j';ij} &=& \sum_\alpha (D_\alpha)_{i'i} (D_\alpha)_{j'j}^{*} \nonumber\\
\mathfrak{B}_{i'i;j'j} &=& \sum_\alpha (D_\alpha)_{i'j'} (D_\alpha)_{ij}^{*} 
\end{eqnarray}
Note that in the last line above, we used Eq.~(\ref{Eq.rearrange}) where $\mathfrak{B}$ was introduced via reshuffling of indices.

\subsection{Stinespring's system-environment representation}

Given a completely positive, trace-prserving map, $\mathcal{E}(\rho)$~on~$\mathcal{B}(\mathcal{H}_{S})$, there exists a space $\mathcal{H}_E$ and a state $|0\rangle \in \mathcal{H}_E$ such that 
\begin{equation}
\mathcal{E}(\rho) = \text{tr}_{E} \Big(U(t) [\rho_{S}(0)\otimes|0\rangle\langle 0|] U^{\dagger}(t)\Big),
\label{stinespringunitary}
\end{equation} 
where $U(t)$ is a unitary on $\mathcal{H}_{S}\otimes \mathcal{H}_{E}$. Eq.~(\ref{stinespringunitary}) is known as Stinespring's system-environment representation of the map \cite{stinespring55}.
Given the Kraus representation of the map on $\mathcal{H}_S$, with Kraus operators $D_{\alpha}$ and considering an orthonormal basis $\{|\alpha\rangle\}$ in $\mathcal{H}_{E}$, we define the unitary $U$ as 
\begin{equation}
U |\phi\rangle_S \otimes |0\rangle_E \equiv \sum_{\alpha} D_{\alpha} |\phi\rangle_S \otimes |\alpha\rangle_E,
\label{isometry}
\end{equation} 
where $|\phi\rangle_S$ is a pure state on $\mathcal{H}_S$ and $U$ is on the space $\mathcal{H}_S \otimes \mathcal{H}_E$. Plugging Eq.~(\ref{isometry}) into Eq.~(\ref{stinespringunitary}) recovers the Kraus form of the map, as in Eq.~(\ref{krausrep}). 

 Note that this was what we had shown in the beginning, in Eq.~(\ref{envrep1}) and Eq.~(\ref{parttrace1}). This representation therefore gives the prescription to construct a unitary from the Kraus operator elements, by dilating the Hilbert space. This is therefore known as Stinespring's dilation theorem. If the dimension of input and output spaces are not the same, then the unitary is replaced by an isometry. We can construct a unitary from the isometry by suitable increase of dimensions of the environmental Hilbert space, $\mathcal{H}_E$. 
 
 An isometry $V$ is the one which satisfies the one-way relation 
$V V^{\dagger} = \mathbb{1}$ unlike a unitary $U$ which satisfies $U U^{\dagger} = \mathbb{1} = U^{\dagger} U$. An isometry therefore is not a square matrix, since it maps two spaces of different dimensions.

\subsection{Rank of $\mathfrak{B}$ and the number of Kraus matrices}

As we have seen, the Kraus matrices are nothing but the matricized versions of the eigenvectors of $\mathfrak{B}$. The Kraus matrices obtained by diagonalizing $\mathfrak{B}$ are usually referred to as the canonical set of Kraus operators. The number of canonical Kraus matrices therefore is set by the dimension of the system under consideration. For an $n$~dimensional system, the $\mathfrak{B}$~matrix is $n^2$ dimensional and hence the maximum rank, which is equal to the number of non zero eigenvalues it can have is $n^2$. Eq.~(\ref{parttrace1}) and Eq.~(\ref{krausrep2}) give us the impression that the number of Kraus matrices is set by the dimensions of the environment. We have seen that the Kraus representation is defined up to a unitary and hence one can construct a new set of Kraus operators $\{K_\beta\}$ (more than $n^2$ in number) using a suitable unitary of appropriate dimensions, such that $ K_\beta  = U_{\beta\alpha} D_\alpha $. Hence, the rank of the map is the number of non-zero eigenvalues of the dynamical matrix $\mathfrak{B}$.

\subsection{Choi's theorem on completely positive maps and the $\mathfrak{B}$~matrix in disguise}

Choi \cite{choi1975} showed that a linear map $\mathcal{E}: \mathcal{B}(\mathcal{H}_{S}) \to\mathcal{B}(\mathcal{H}_{S'})$ is completely positive if $ C_{\mathcal{E}}:= \mathcal{E} \otimes \mathbb{1} |I\rangle\rangle \langle\langle I|$ is positive. The matrix $ C_{\mathcal{E}}$ is often called the Choi matrix. We show that the Choi matrix and the $\mathfrak{B}$~matrix are the same. Since $\mathcal{E}: \mathcal{B}(\mathcal{H}_{S}) \to\mathcal{B}(\mathcal{H}_{S'})$ and $\mathcal{E}(\rho) = \sum_\alpha D_{\alpha}^{}\rho D_{\alpha}^\dagger$. Using Eq.~(\ref{identity3}) and Eq.~(\ref{identity1}) in succession,
\begin{eqnarray}
\mathcal{E}(\rho) &=& \text{tr}_{\mathcal{H}_{S}} \Big ( \sum_\alpha |D_{\alpha}\rho\rangle\rangle  \langle\langle D_{\alpha}|\Big) \nonumber\\
&=& \text{tr}_{\mathcal{H}_{S}} \Big (\mathbb{1}\otimes \rho^{T} \sum_\alpha |D_{\alpha}\rangle\rangle  \langle\langle D_{\alpha}|\Big).
\label{bchoiconn}
\end{eqnarray} 
Now, Eq.~(\ref{identity2}) tells us that $|D_{\alpha}\rangle\rangle = D_{\alpha} \otimes \mathbb{1} |I\rangle\rangle \langle\langle I|$. Hence, Eq.~(\ref{bchoiconn}) can be written as
 \begin{equation}
\label{bchoiconn2}
 \mathcal{E}(\rho) = \text{tr}_{\mathcal{H}_{S}} \Big (\mathbb{1}\otimes \rho^{T} \sum_{\alpha} D_{\alpha}\otimes \mathbb{1} |I\rangle\rangle \langle\langle I|  D_{\alpha}^{\dagger} \otimes \mathbb{1} \Big).
\end{equation}
But $\sum_{\alpha} D_{\alpha}\otimes \mathbb{1} |I\rangle\rangle \langle\langle I|  D_{\alpha}^{\dagger} \otimes \mathbb{1}$ is the Choi matrix, $C_{\mathcal{E}}$. Eq.~(\ref{bchoiconn}) and Eq.~(\ref{bchoiconn2}) reveal the fact that the Choi matrix and $\mathfrak{B}$ are one and the same. Thus, one can write down the action of the map in the $\mathfrak{B}$ form as 
 \begin{eqnarray}
\label{baction}
 \mathcal{E}(\rho) &=& \text{tr}_{\mathcal{H}_{S}} \Big (\mathbb{1}\otimes \rho^{T} \mathfrak{B} \Big),\nonumber\\
 &=& \text{tr}_{\mathcal{H}_{S}} \Big (\mathbb{1}\otimes \rho^{T} C_{\mathcal{E}} \Big).
\end{eqnarray}
It becomes transparent that $\mathfrak{B}$ has in it encoded the definition of complete positivity.  One can now prove the Choi's theorem easily. Using Eq.~(\ref{baction}), by going backwards in the steps outlined above, the form of the map boils down to Eq.~(\ref{eqn:krausform}), which was shown to be CP in Eq.~(\ref{eqn:krauscpcondn}).

For completeness, let us look into the $\chi$ matrix as well.

The $\chi$ matrix can be written as \begin{equation} 
\label{chidef}
\chi =  \sum_{\alpha = 1}^{r} \widetilde{D}_{\alpha} \widetilde{D}^{T}_{\alpha}
\end{equation}where
\begin{equation*}
\widetilde{D}_{1}=\begin{bmatrix}
D_{1}^{(1)} \\
D_{1}^{(2)}   \\
\vdots \\
D_{1}^{(d)}   
\end{bmatrix}, \text{with}\quad D_{1}^{(d)} = \text{Tr}(D_{1}A_{d})   \end{equation*} $\{A_d\}$ denotes the basis used in evaluating the $\chi$ matrix.

Similarly,
\begin{equation*}
\widetilde{D}^{T}_{1}=\begin{bmatrix}
d_{1}^{(1)}\cdots d_{1}^{(d)}\\  
\end{bmatrix}, \text{where}\quad d_{1}^{(d)} = \text{Tr}(D^{T}_{1}A_{d})   \end{equation*}

Now, one can easily recognize that the $\chi$ matrix introduced in Eq.~(\ref{eqn:chikraus}) and the $\mathfrak{B}$ are the same, if the basis chosen for the expansion of Kraus operators is $\{A_d\} = \{E_{ij} =  |e_i \rangle \langle e_j|\}$. 

\subsection{Trace preservation conditions}

Having seen the different representations of the map and complete positivity, we now see how the conditions of Trace Preservation (TP) are encoded in the various representations.

We have already seen from Eq.~(\ref{eqn:krausform1}) that if the map is in the Kraus form, 
\begin{equation*}
 \sum_{\alpha} D_\alpha^\dagger D_\alpha^{\vphantom{\dagger}} = \mathbb{1},
		\end{equation*} 
then the corresponding condition for the $\mathfrak{B}$ form is
\begin{equation}
  \text{tr}_{\mathcal{H}_{S^{\prime}}} (\mathfrak{B}) = \mathbb{1}_S.
\end{equation}
This can be easily obtained by identifying $\mathfrak{B}$ with the definition of the Choi matrix and noting its action as shown in Eq.~(\ref{baction}).
Also, the trace of $\mathfrak{B}$ and hence the sum of its eigenvalues is equal to $N$, where $N$ is the dimension of the system state space on which the map acts. 
\begin{eqnarray}
  \text{tr}(\mathfrak{B}) &=&   \text{tr} \Big(\sum_{i,j = 1}^{N} \mathcal{E}(|e_i \rangle \langle e_j|) \otimes |e_i \rangle \langle e_j|\Big)\nonumber\\
  &=&   \text{tr} \Big(\sum_{i = 1}^{N} \mathcal{E}(|e_i \rangle \langle e_i|) \Big)\nonumber\\
  &=& N.
\end{eqnarray}
The $\mathfrak{B}$~matrix satisfies all the properties of a valid density matrix, apart from its trace. $\frac{1}{N}\mathfrak{B}$ becomes a valid state and this correspondence between states and dynamical matrices is referred to as the Choi--Jamiolkowski isomorphism. Thus, quantum processes (maps) can be easily mapped to states which makes the analysis of Quantum Process Tomography \cite{chuang_prescription_1997} easier. Using the details presented in Section~(\ref{abindexkraus}), it is left as an exercise to the reader to show the condition of trace preservation in the $\mathcal{A}$ form as outlined in Eq.~(\ref{arest}).

\subsection{Merits of various representations}

Let us now take a look at the merits of various representations. 

$\mathcal{A}$ form: To represent the action of a map, i.e., the output of a map acting on a state, this form can be used, if one writes the density matrix as a column matrix. Checking for CP and TP, the $\mathcal{A}$ form is not the best choice.

$\mathfrak{B}$ form: For checking whether a map is completely positive, it amounts to evaluating the positivity of $\mathfrak{B}$. For checking any property related to the map itself, the $\mathfrak{B}$~form is the best choice. Checking for TP in the $\mathfrak{B}$ form amounts to evaluating a partial trace and this is a good choice. 

Kraus form: To represent the action of a map, the Kraus representation is the usual choice. The Kraus form implies CP and conditions for TP is also straightforward in this form.

{$\chi$ matrix: A positive $\chi$ is an indicator of the CP nature of the map. In Quantum Process Tomography, the $\chi$~matrix representation is used usually. For example, for maps on a single qubit, the usual basis of representing the $\chi$ matrix is the basis of Pauli matrices along with the identity matrix. The $\chi$ matrix and the $\mathfrak{B}$~matrix are the same if the basis is the standard basis.}

Stinespring's form: Physical realizations of maps in the laboratory can be tried out using extended system-environment unitaries which stem from the Kraus representation followed by Stinespring dilation.

\section{Allowed operations on a qubit: balls, spindles and pancakes}
\label{qubitmaps}

The requirement of complete positivity for the reduced dynamics places restrictions on the allowed behavior of an open quantum system. Let us look into the restrictions where our system $S$ is a qubit.
The state of the qubit is represented by a $2 \times 2$, Hermitian, positive, complex matrix $\rho$ of unit trace. The density matrix is\begin{equation}
  \label{Eq.1q1}
  \rho = \frac{1}{2}(\mathbb{1} + a_i \sigma_i) = \frac{1}{2} \left( \begin{array}{cc}1 + a_3 & a_1 - \imath a_2\\ a_1 + i a_2  & 1-a_3 \end{array} \right),
\end{equation}
where $\sigma_i$ are the Pauli matrices.
\[ \sigma_1 = \left( \begin{array}{cc}0 & 1 \\ 1 & 0 \end{array} \right)  ; \quad   \sigma_2 = \left( \begin{array}{cc}0 & -\imath \\ i & 0 \end{array} \right)  ; \quad \sigma_3 = \left( \begin{array}{cc}1 & 0 \\ 0  & -1 \end{array} \right). \]
The vector $\vec{a} = (a_1 \,,\, a_2 \,,\, a_3)$ is called the Bloch vector and physical states of the qubit correspond to $|\vec{a}| \leq 1$. The space of all one qubit states can therefore be viewed as all the points on or inside the \emph{Bloch ball}, which is the unit ball in the space spanned by $a_1$, $a_2$ and $a_3$. 

\subsection{Unital maps}

Let us consider a map~$\mathcal{E}$  on the basis $\{\mathbb{1}, \sigma_{i = 1, 2, 3}\}$ defined as follows,
\begin{eqnarray}
\mathcal{E}(\mathbb{1}) &=& \mathbb{1}, \nonumber\\
\mathcal{E}(\sigma_{i}) &=& z_{i} \sigma_{i}, z_i \in \mathbb{R}.
\label{unitaldef}
\end{eqnarray}
This means that the map on $\rho$ scales each of the three independent directions as follows:
\begin{equation}
  \label{Eq.1q2}
  \rho' = \mathcal{A}_{\text{U}} \rho = \frac{1}{2}(\mathbb{1}+ z_i a_i \sigma_i) \quad {\mbox{with}} \quad 0 \leq  |z_i|  \leq 1.
\end{equation}
Rearranging the matrices $\rho$ and $\rho'$ and constructing the following equation 
\begin{equation}
  \label{Eq.1q3}
 \frac{1}{2} \left( \begin{array}{c} 1+z_3a_3 \\ z_1a_1-\imath z_2a_2 \\ z_1a_1+\imath z_2a_2 \\ 1-z_3a_3 \end{array} \right) = \mathcal{A}_{\text{U}}  \cdot \frac{1}{2}\left( \begin{array}{c} 1+a_3 \\ a_1-\imath a_2 \\ a_1+\imath a_2 \\ 1-a_3 \end{array} \right),
\end{equation}
the linear operator $\mathcal{A}_{\text{U}}$ may be written down by inspection to be
\begin{equation}
  \label{Eq.1q4}
  \mathcal{A}_{\text{U}} = \frac{1}{2}\left( \begin{array}{cccc}
1+z_3 & 0 & 0 & 1-z_3 \\ 0 & z_1 + z_2 & z_1 -z_2 & 0 \\ 0 & z_1 - z_2 & z_1 + z_2 & 0 \\ 1-z_3 & 0 & 0 & 1+z_3
 \end{array} \right).
\end{equation}
From the $\mathcal{A}_{\text{U}}$ matrix we obtain the dynamical matrix as
\begin{equation}
  \label{Eq.1q5}
   \mathfrak{B}_{\text{U}}= \frac{1}{2}\left( \begin{array}{cccc}
1+z_3 & 0 & 0 & z_1+z_2 \\ 0 & 1-z_3 & z_1 -z_2 & 0 \\ 0 & z_1 - z_2 & 1-z_3 & 0 \\ z_1+z_2 & 0 & 0 & 1+z_3
 \end{array} \right).
\end{equation}
The eigenvalues of $\mathfrak{B}_{\text{U}}$ are
\begin{eqnarray}
  \label{Eq.1q6}
  \lambda_1 & = & \frac{1}{2} ( 1 + z_1 - z_2 - z_3), \nonumber \\
   \lambda_2 & = & \frac{1}{2} ( 1 - z_1 + z_2 - z_3), \nonumber \\
    \lambda_3 & = & \frac{1}{2} ( 1 - z_1 - z_2 + z_3), \nonumber \\
     \lambda_4 & = & \frac{1}{2} ( 1 + z_1 + z_2 + z_3).
\end{eqnarray}
If $\mathfrak{B}_{\text{U}}$ has to be completely positive then all $\lambda_i$ must be positive semi-definite. This means that the scaling parameters $z_i$ have to be such that
\begin{equation}
  \label{Eq.1q7}
  z_3 \leq 1 - (z_2 - z_1) \quad {\mbox{given that}} \quad z_1 \leq z_2.
\end{equation}

Eq.~(\ref{Eq.1q7}) tells us that complete positivity of the dynamical map put restrictions on the possible transformations of the Bloch ball. As we can see, the map cannot take the unit ball into a unit-pancake (the map which projects the states on the Bloch ball to any plane) because this would correspond to the choice $z_1=0$ and $z_2 = z_3 =1$ which violates (\ref{Eq.1q7}). This means that a relaxation along only one of the three orthogonal directions is not allowed by the constraint of complete positivity. 

In other words, the transformations can produce balls of shorter radii and ellipsoids, but pancakes of unit radius are not allowed.  

Maps when acted on $\mathbb{1}$ leave them unchanged are called {\em unital}. For qubits, such maps do not move the center of the Bloch ball. Since $\mathcal{E}(\mathbb{1}) = \mathbb{1}$, 
\begin{eqnarray}
  \label{unitalcondnkrausb}
 \sum_{\alpha} D_\alpha D_\alpha^\dagger = \mathbb{1},\nonumber\\
  \text{tr}_{\mathcal{H}_{S}} (\mathfrak{B}) = \mathbb{1}_{S^{\prime}}.
  \end{eqnarray}

\begin{figure}[t!]
\includegraphics[width=85mm]{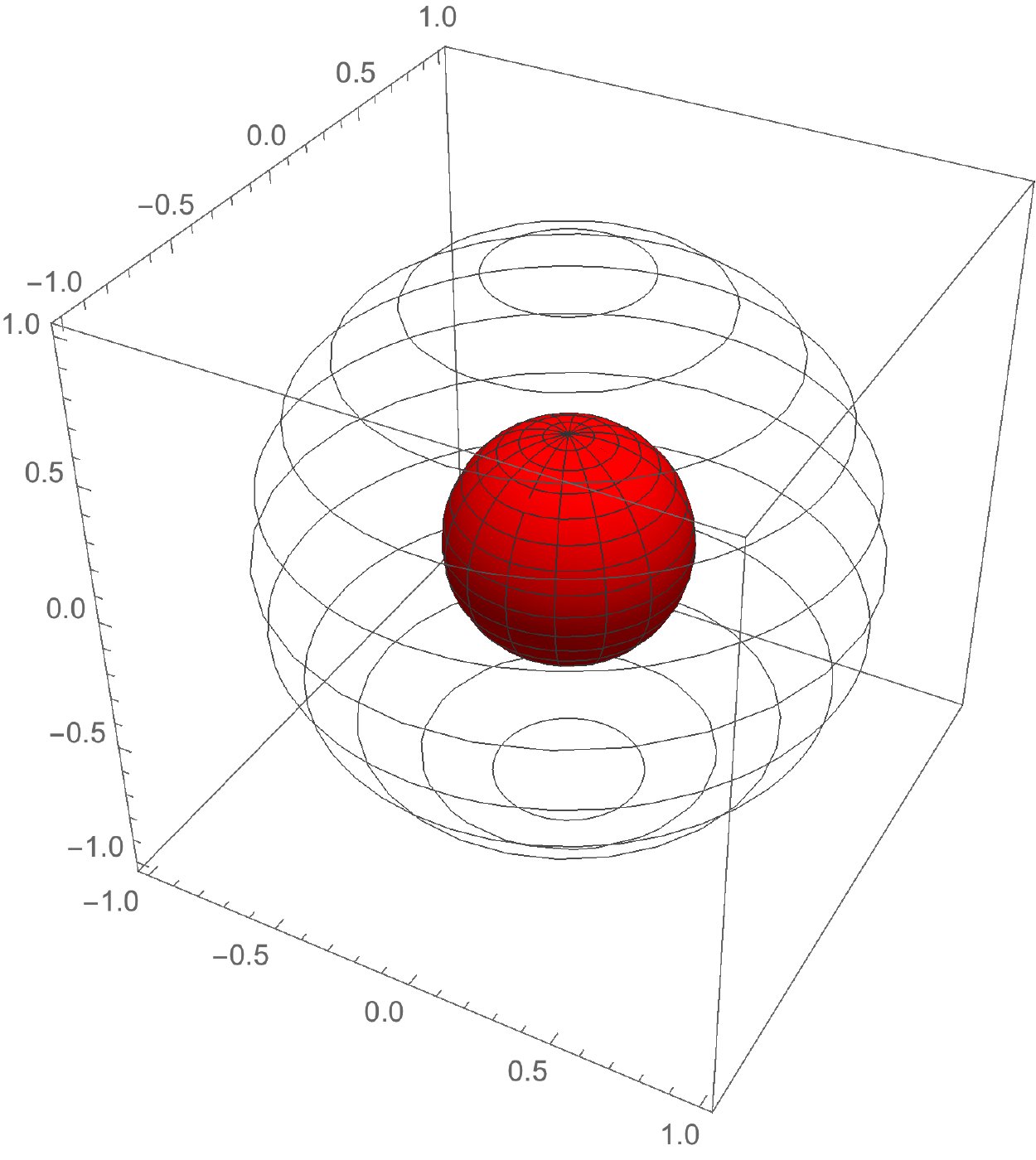}
\caption{Action of a depolarizing channel on the Bloch ball. The Bloch ball (outer ball of unit radius) is shown in white. The image is plotted in red.  \label{fig:1}}
\end{figure}

\subsection{Non-unital maps}

 We can also have dynamical maps with affine shifts that move the center of the Bloch ball as well. Such maps are called {\em non-unital}.  They are obtained if the action of the map~$\mathcal{E}$  on the basis $\{\mathbb{1}, \sigma_{i = 1, 2, 3}\}$ is as follows.
\begin{eqnarray}
\mathcal{E}(\mathbb{1}) &=& \mathbb{1} + \sum_{i = 1}^{3} t_{i} \sigma_{i},\nonumber\\
\mathcal{E}(\sigma_{i}) &=& z_{i} \sigma_{i}.
\label{nunitaldef}
\end{eqnarray}
From this, one can easily see that the general form of the dynamical matrix,$\mathfrak{B}_{\text{NU}}$ representing a trace-preserving $\mathcal{E}$ on a qubit which is non-unital is given as 
\begin{equation}
\label{eqmap}
 \frac{1}{2} \left(
\begin{array}{cccc}
1+ t_3+z_3 & t_1-\imath t_2 & 0 & z_1+z_2 \\
 t_1+\imath t_2 & 1-t_3-z_3 & z_1-z_2 & 0 \\
 0 & z_1-z_2 & 1+t_3-z_3 & t_1-\imath t_2 \\
 z_1+z_2 & 0 & t_1+\imath t_2 & 1-t_3+z_3 \\
\end{array}
\right).
\end{equation}
The conditions for complete positivity for non-unital maps on a qubit are a bit complicated  \cite{beth_ruskai_analysis_2002}, which we do not give here.

\subsection{Representation of maps on a qubit in terms of the Bloch vector}

There is a compact way of representing the output of a map on a qubit, in terms of the Bloch vector. From Eq.~(\ref{Eq.1q1}), we know that the density matrix of a qubit is $\rho = \frac{1}{2}(\mathbb{1} + \vec{a}.\vec {\sigma})$. From Eqs: (\ref{unitaldef}) and (\ref{nunitaldef}), we can write the action of any CP map on a qubit in a compact form as
\begin{equation}
\mathcal{E}\Big(\frac{1}{2}\mathbb{1} + \vec{a}\cdot\vec{\sigma}\Big) = \frac{1}{2}\mathbb{1} + \Big ( \frac{1}{2} \vec{t} + T\vec{a}\Big). \vec{\sigma},
\label{actblochvec}
\end{equation}
where 
 \begin{equation}
T = \left(
\begin{array}{ccc}
 x_1 & 0 & 0 \\
 0 & x_2 & 0 \\
 0 & 0 & x_3 \\
\end{array}
\right),
\label{tdef}
\end{equation}
and 
\begin{equation}
\vec{t} = 
\left(
\begin{array}{c}
 t_1 \\
 t_2 \\
 t_3 \\
\end{array}
\right).
\label{tvecdef}
\end{equation}
If $\vec{t} = \vec{0}$, then the map is unital. Non-unital maps therefore have a non-zero $\vec{t}$. 

\subsection{Examples of a few standard maps on the qubit}

\begin{figure}[t!]
\includegraphics[width=85mm]{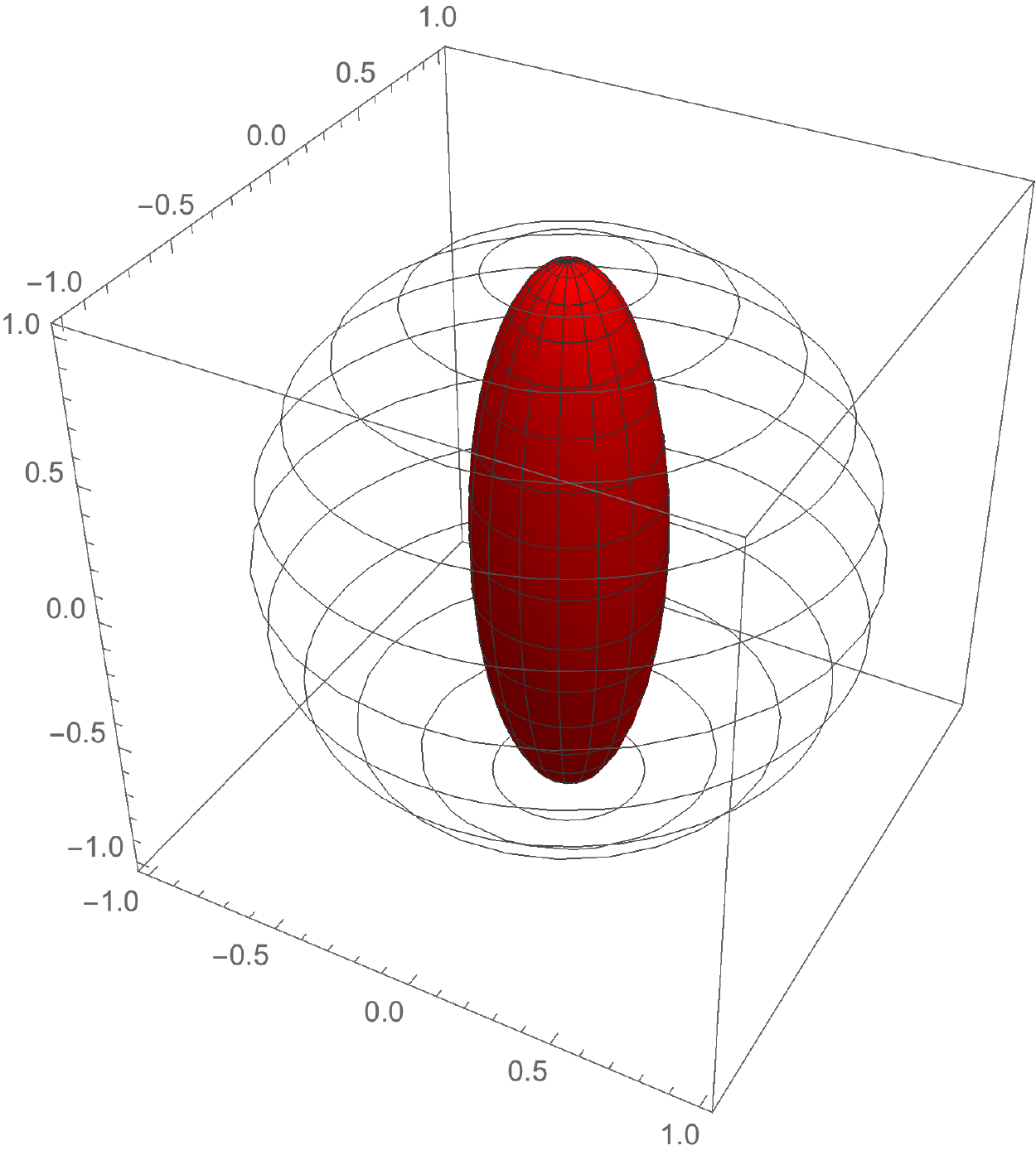}
\caption{Action of a phase damping channel on the Bloch ball. The Bloch ball (outer ball of unit radius) is shown in white. The image is plotted in red. \label{fig:2}}
\end{figure}

We illustrate the examples of a few standard maps acting on a qubit.  
We start with a density matrix $\rho_0$ and the corresponding Bloch vector $\vec{a}$ is given by
\begin{equation}
  \rho_0 =  \left( \begin{array}{cc}\rho_{00} & \rho_{01}\\ \rho_{10}  & \rho_{11} \end{array} \right), \quad\text{and}\quad \vec{a} = \begin{pmatrix} \rho_{01}+\rho_{10} \\ i( \rho_{01} +  \rho_{10}) \\  \rho_{00}-\rho_{11} \end{pmatrix}.
\end{equation}
The three components of the Bloch vector are the traces of the density matrix with the three Pauli matrices, namely, $a_i = \mathrm{tr} (\rho\sigma_i)$.
In Figures \ref{fig:1},\ref{fig:2},\ref{fig:3},\ref{fig:4},\ref{fig:5},\ref{fig:6}, we see how the Bloch ball (outer ball of unit radius, shown in white) gets mapped under the action of various standard maps.

\subsubsection{Depolarizing channel}

If the action of the map is such that the entire Bloch ball is contracted by a constant factor, the process is called depolarizing. This drives a qubit to the maximally mixed state, $\mathbb{1}/2$, which means that the action of the map is to drive the qubit to the centre of the Bloch ball. The Kraus operators for this process are\begin{eqnarray}
  D_1 = \sqrt{1 - \frac{3p}{4}}\mathbb{1}\quad ,\quad  D_2 = \sqrt{ \frac{p}{4}}\sigma_{1},\nonumber\\
  D_3 = \sqrt{\frac{p}{4}}\sigma_{2}\quad ,\quad  D_4 = \sqrt{ \frac{p}{4}}\sigma_{3}.
  \end{eqnarray}
  Under this process, the Bloch vector changes to
  \begin{equation}
\vec{a} =  (1 - p)  \begin{pmatrix} \rho_{01}+\rho_{10} \\ \imath( \rho_{01} +  \rho_{10}) \\  \rho_{00}-\rho_{11} \end{pmatrix}.
\end{equation}
Here, the Bloch ball gets mapped to a ball of shorter radius, see Figure \ref{fig:1}.

\subsubsection{Phase damping channel}

Under the phase damping channel, the Bloch ball is contracted to a prolate spheroid about one of the axes. Here, we show the $z$ axis. The Kraus operators for the phase damping process are\begin{equation}
  D_1 = \sqrt{1 - \frac{p}{2}}\mathbb{1}\quad ,\quad  D_2 = \sqrt{ \frac{p}{2}}\sigma_{3}.  \end{equation}
  Under this process, the Bloch vector evolves to
  \begin{equation}
\vec{a} = \begin{pmatrix} (1 - p)(\rho_{01}+\rho_{10}) \\ \imath(1 - p)( \rho_{01} +  \rho_{10}) \\  \rho_{00}-\rho_{11} \end{pmatrix}.
\end{equation}
Here, the map compresses the Bloch ball to an ellipsoid. This indicates that the action of the map is to drive a superposed state to a statistical mixture, see Figure \ref{fig:2}.

\begin{figure}[t!]
\includegraphics[width=85mm]{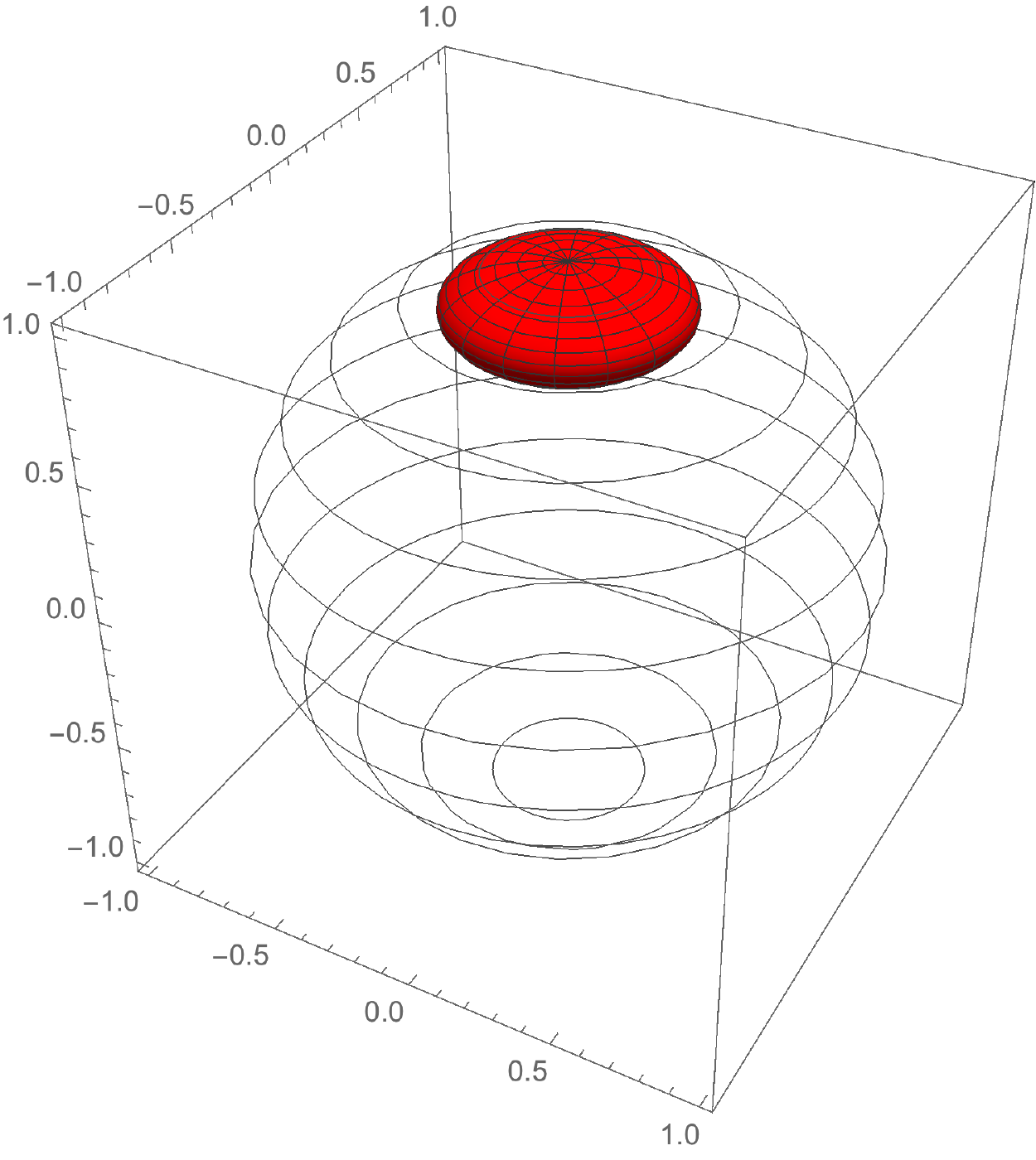}
\caption{Action of an amplitude damping channel on the Bloch ball. The Bloch ball (outer ball of unit radius) is shown in white. The image is plotted in red.\label{fig:3}}
\end{figure}

\subsubsection{Amplitude damping channel}

This is an example for a map which is non-unital. Here, the entire Bloch ball is shrunk to one of the poles of the Bloch ball.  The Kraus operators are\begin{equation}
  D_1 = \left( \begin{array}{cc}1 & 0 \\ 0 &\sqrt{1 - p} \end{array} \right) , \quad   D_2 = \left( \begin{array}{cc}0&\sqrt{p} \\ 0&0 \end{array}\right). \end{equation}
  Under this process, the Bloch vector becomes
  \begin{equation}
 \vec{a} = \begin{pmatrix} \sqrt{1 - p}(\rho_{01}+\rho_{10}) \\ \imath \sqrt{1 - p}( \rho_{01} +  \rho_{10}) \\  \rho_{00}-(1 - 2p)\rho_{11} \end{pmatrix}.
\end{equation}
The map being non-unital, not only compresses the Bloch ball, but shifts the centre as well, see Figure \ref{fig:3}.

\subsubsection{Unitary evolution}

\begin{figure}[t!]
\includegraphics[width=85mm]{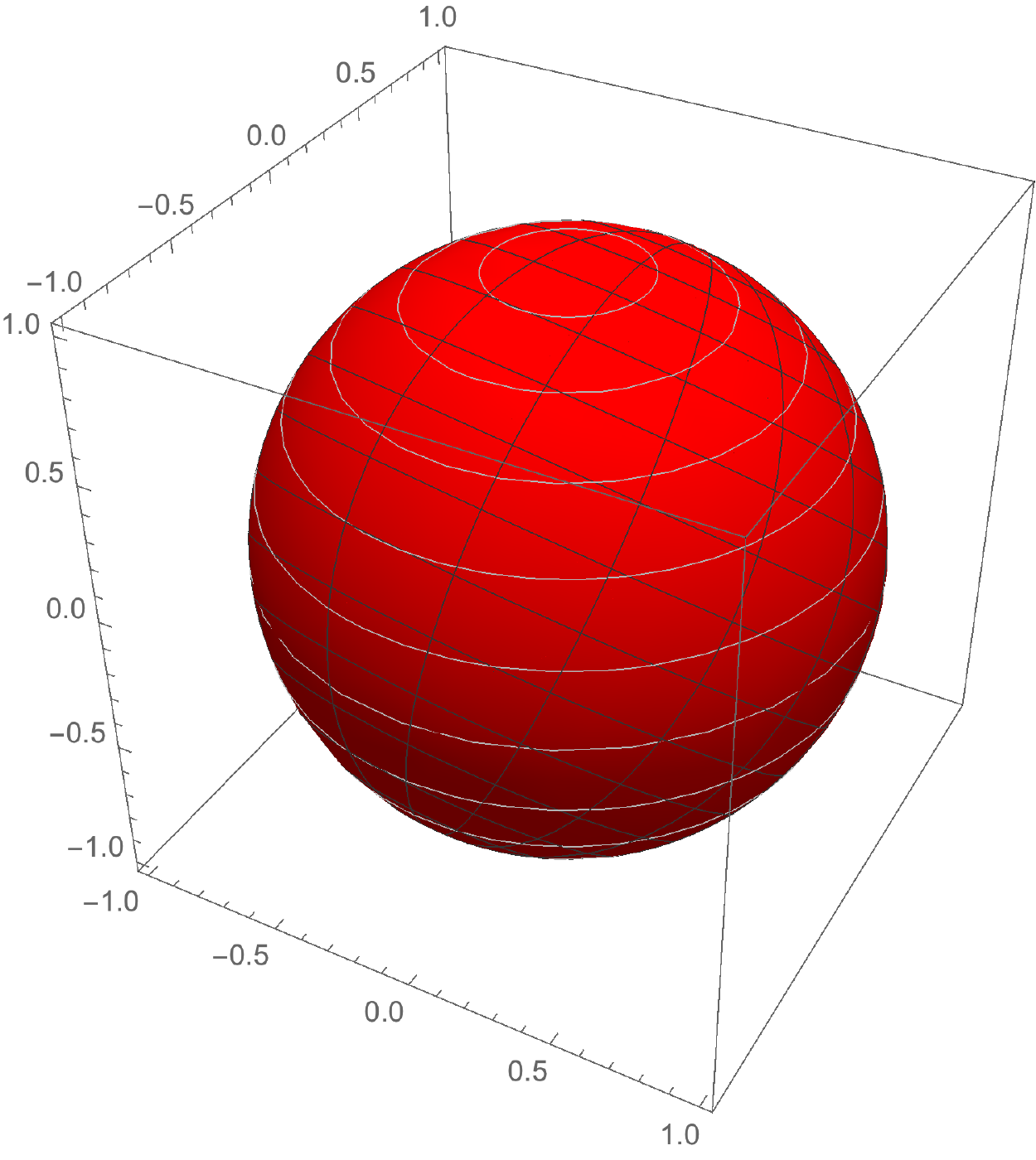}
\caption{Action of unitary on the Bloch ball. The Bloch ball gets mapped to itself. The image is plotted in red. \label{fig:4}}
\end{figure}

Unitary evolutions rotate the Bloch-ball and therefore the entire Bloch ball gets mapped to itself, see Figure \ref{fig:4}.

\section{Not completely positive maps}
\label{ncpmaps}

We have addressed CP maps so far. However, there are maps which are known to violate complete positivity. In the pioneering paper of Sudarshan et al. \cite{sudarshan61a} and other succeeding ones \cite{jordan_dynamical_1961,jordan_dynamical_1962} complete positivity was not assumed, but was focused only on characterizing the most general dynamical framework for quantum systems in terms of linear maps acting on density matrices. However, later on, it became quite acceptable in the community that complete positivity should be imposed in open quantum evolution. However, there have been active debates over this issue. For instance, Simmons and Park argued that CP maps are incompatible with the phenomenological theory of spin relaxation \cite{simmons_completely_1981,raggio_remarks_1982,simmons_another_1982}.There has also been an active exchange between Pechukas and Alicki\cite{pechukas_reduced_1994,alicki_comment_1995,pechukas_pechukas_1995}. For initially correlated system and environment, it was shown that the reduced dynamics can be \emph{not completely positive} (NCP) \cite{jordan_dynamics_2004}. This also led to further activity along these lines, especially with applications to quantum error-correction \cite{shabani_maps_2009}. Very recently, it was shown that CP is a necessary and sufficient condition for the data-processing inequality to be valid\cite{buscemi_complete_2014}. 

\subsection{Operator sum-difference representation of NCP maps}

We have seen that complete positivity is implied by the fact that all eigenvalues of the dynamical matrix $\mathfrak{B}$ are positive, which leads to the Operator Sum Representation, Eq.~(\ref{krausrep}). If $\mathfrak{B}$ has one or more negative eigenvalues, then the map is NCP. Let us consider $\mathfrak{B}$ of dimensions $N^2$, which means it can have a maximum of $N^2$ non-zero eigenvalues. Letting $\lambda_\alpha$\,\! and $|\Lambda^{(\alpha)}\rangle$\,\! to be the corresponding eigenvalues and eigenvectors respectively of $\mathfrak{B}$ and assuming that only the eigenvalues from $\alpha = 1,2...k$ are positive, then we define $D_{\alpha} = \sqrt{\lambda_{\alpha}}\thinspace \mathrm{mat}  |\Lambda^{(\alpha)}\rangle$ and $F_{\alpha} = |\sqrt{\lambda_{\alpha}}\thinspace |\thinspace \mathrm{mat}  |\Lambda^{(\alpha)}\rangle$. This leads to the Operator Sum-Difference Representation,
 \begin{equation}
  \label{osddecomp}
	\rho^\prime = \sum_{\alpha = 1}^{k} D_{\alpha}^{}\rho D_{\alpha}^\dagger - \sum_{\alpha = k+1}^{N^2} F_{\alpha}^{}\rho F_{\alpha}^\dagger. 
	\end{equation}
For the negative eigenvalues, one takes the modulus, $|\sqrt{\lambda_{\alpha}}\thinspace|$ to write the Kraus operators. The condition for trace preservation becomes
\begin{equation}
 \sum_{\alpha= 1}^{k} D_\alpha^\dagger D_\alpha -  \sum_{\alpha = k+1}^{N^2} F_\alpha^\dagger F_\alpha = \mathbb{1}.
\end{equation}

\subsection{Examples of NCP maps}

The study of NCP maps are very interesting, but we do not intend to discuss them in detail in this article. Next, we give a few examples of maps on a qubit which are NCP. 

\subsubsection{Spin-reversal}

Consider a map on a qubit which flips the Bloch vector $\vec{a} \to -\vec{a}$. This means that \begin{eqnarray}
\mathcal{E}(\mathbb{1}) &=& \mathbb{1}, \nonumber\\
\mathcal{E}(\sigma_{i}) &=& - \sigma_{i}.
\end{eqnarray}The corresponding dynamical map can be evaluated to be
\begin{equation}
\mathfrak{B}_{\text{SR}} = \left(
\begin{array}{cccc}
 0 & 0 & 0 & -1 \\
 0 & 1 & 0 & 0 \\
 0 & 0 & 1 & 0 \\
 -1 & 0 & 0 & 0 \\
\end{array}
\right).
\end{equation}
$\mathfrak{B}_{\text{SR}}$ has one of its eigenvalues as -1 indicating that the map is not CP. This tells us why a universal NOT gate {does not correspond to a CP map.}

\subsubsection{Transpose map}

Consider the transpose map on a qubit. 
$\mathcal{E}(\rho) = \rho^{T}$. The dynamical matrix associated with this operation is
\begin{equation}
\mathfrak{B}_{\text{T}} = \left(
\begin{array}{cccc}
 1 & 0 & 0 & 0 \\
 0 & 0 & 1 & 0 \\
 0 & 1 & 0 & 0 \\
 0 & 0 & 0 &1 \\
\end{array}
\right),
\end{equation}
which has a negative eigenvalue indicative of the NCP nature of the map.

\subsubsection{Projection of the Bloch ball onto the unit plane}

\begin{figure}[t!]
\includegraphics[width=85mm]{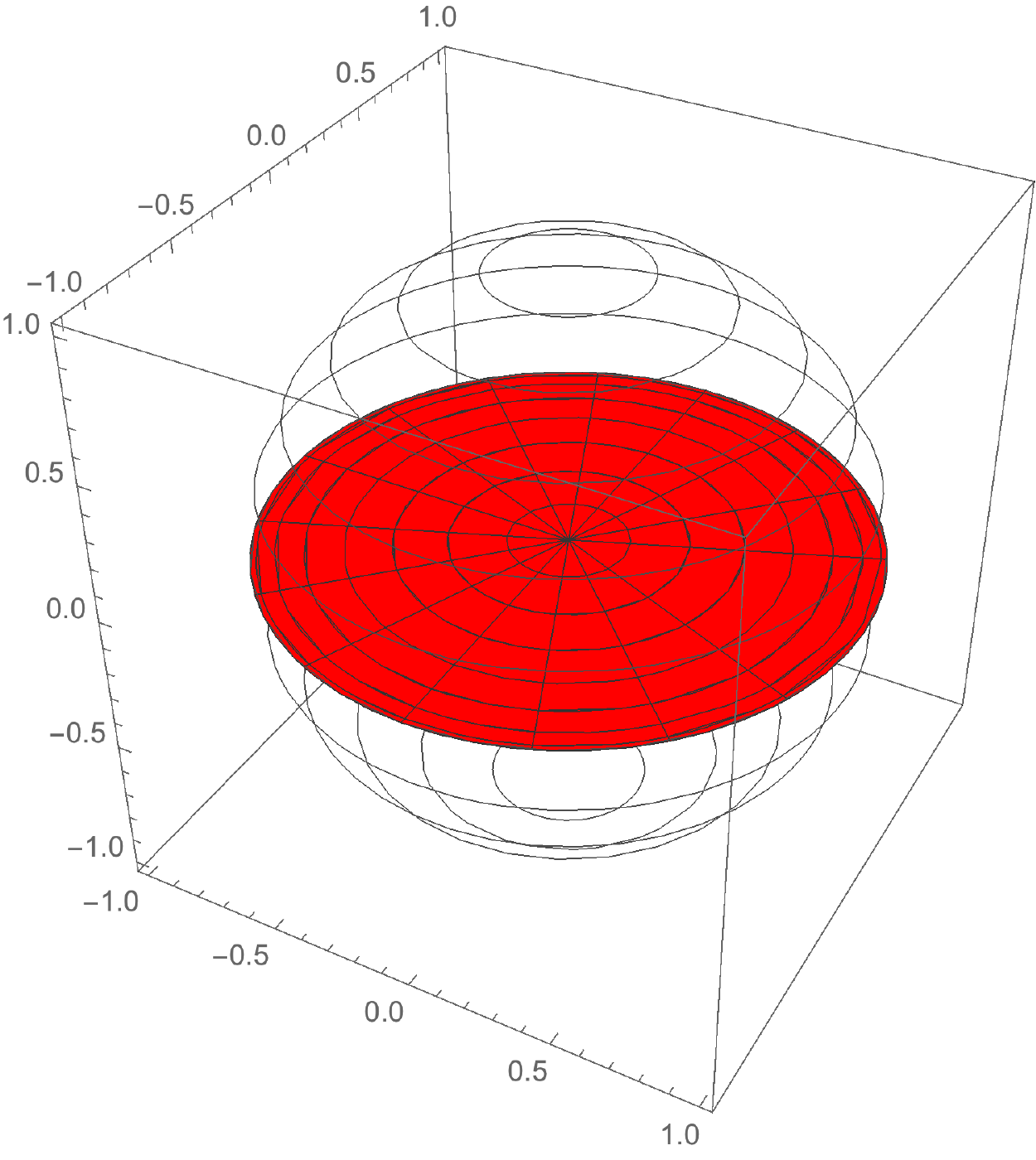}
\caption{Action of the not completely positive (NCP) pancake map. The Bloch ball (outer ball of unit radius) is shown in white. The image is plotted in red.\label{fig:5}}
\end{figure}

The action of the map~$\mathcal{E}$ on the basis $\{\mathbb{1}, \sigma_{i = 1, 2, 3}\}$ is as follows,
\begin{eqnarray}
\mathcal{E}(\mathbb{1}) &=& \mathbb{1}, \nonumber\\
\mathcal{E}(\sigma_{1}) &=&  \sigma_{1},\nonumber\\
\mathcal{E}(\sigma_{2}) &=&  \sigma_{2},\nonumber\\
\mathcal{E}(\sigma_{3}) &=&  0.
\label{pancakencpaction}
\end{eqnarray}
The associated dynamical matrix can be written down by inspection and is
\begin{equation}
\mathfrak{B}_{\text{Proj}} = \left(
\begin{array}{cccc}
 0.5 & 0 & 0 & 1 \\
 0 & 0.5 & 0 & 0 \\
 0 & 0 & 0.5 & 0 \\
 1 & 0 & 0 & 0.5 \\
\end{array}
\right)\end{equation} which has the eigenvalues as 1.5, 0.5. 0.5 and -0.5, which tells that the map is NCP. The physical meaning of Eq.(\ref{pancakencpaction}) is that the action of the map is to project the Bloch-ball to a unit disc in the $x-y$ plane, see Figure~\ref{fig:5}. Therefore, if CP is demanded, such a projection is forbidden.

\begin{figure}[t!]
\includegraphics[width=85mm]{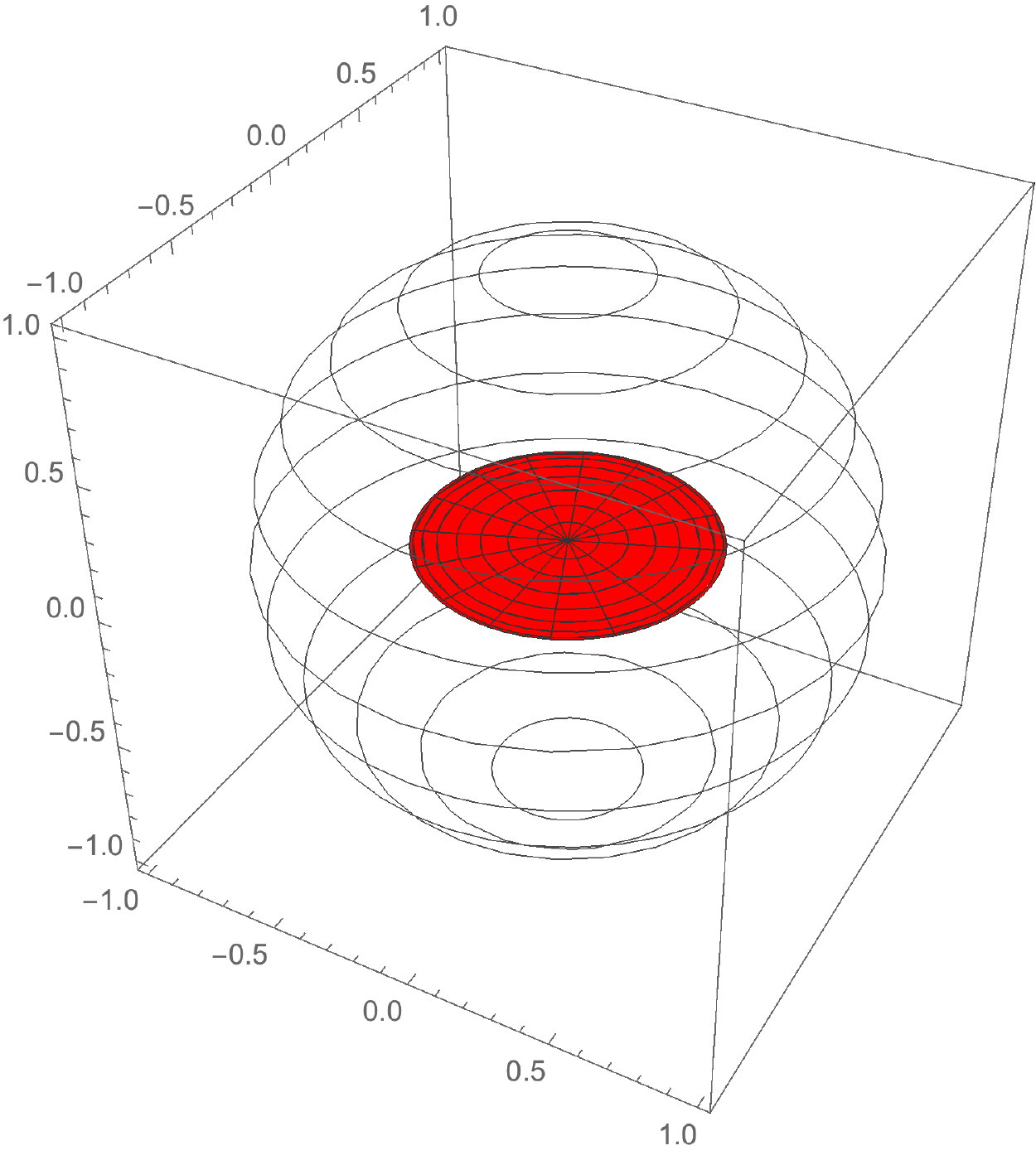}
\caption{Action of the completely positive (CP) pancake map. The Bloch ball (outer ball of unit radius) is shown in white. The image is plotted in red.\label{fig:6}}
\end{figure}

Though a unit disc (one touching the side of the Bloch ball) is restricted by complete positivity, one should notice that the image of Bloch ball can still be a disc in a plane. For instance, if the action of the map is
\begin{eqnarray}
\mathcal{E}(\mathbb{1}) &=& \mathbb{1}, \nonumber\\
\mathcal{E}(\sigma_{1}) &=&  0.5 \sigma_{1},\nonumber\\
\mathcal{E}(\sigma_{2}) &=&  0.5 \sigma_{2},\nonumber\\
\mathcal{E}(\sigma_{3}) &=&  0,
\label{pancakecpaction}
\end{eqnarray}
then the dynamical map is CP and the image is a disc of shorter radius as shown in Figure~\ref{fig:6}.

The reader should not correlate NCP maps with quantum processes that are unphysical. NCP maps should be considered as natural extensions of CP maps albeit the restriction that the domain of action of NCP maps are limited. It must be remembered that whenever a map on a quantum system is spoken of, its action is defined on all possible states of the system. But for NCP maps, this domain gets restricted. That is the action of the map results in valid density matrices as outputs for a subset of states associated with the quantum system. For the case of a qubit, if a map is CP, the entire Bloch ball is a valid domain, but for NCP maps, the positivity domain (the domain of states which leads to valid density matrices) is only a subset and not the entire Bloch ball. The reader can look into \cite{shaji_whos_2005,carteret_dynamics_2008} for more details.

\subsection{Applications of NCP maps}

A bipartite system is entangled if the action of a positive map on one of the subsystems gives a matrix with one or more negative eigenvalues. Therefore, NCP maps are potential candidates for witnessing entanglement. The search for potential entanglement witnesses is an active area of research and the interested reader may refer to \cite{Chruscinski-2014}. The connection between NCP maps and Non-Markovianity has also been looked into and measures of Non-Markovianity based on the NCP nature of the map has been addressed in \cite{rivasprl,rivasreview}. The theory of NCP maps is still an active area of research and potentially holds a wealth of surprises.

\section{Differential form of the map}
\label{kosseq}

We have addressed finite-time maps so far and looked at various representations of the same. For completeness, let us look into the differential form of the map as well. This was done by Gorini, Kossakowski and Sudarshan in \cite{gorini_completely_1976} and by Lindblad \cite{lindblad76}. We shall not follow the mathematically motivated version, but present a heuristic derivation of the same instead. 

Consider an infinitesimally short time-interval, $\Delta t$ during which the state of the system evolves from $\rho$ to $\Delta\rho$. The map is then written as
\begin{equation}
   \label{Eq.koss1}
\rho + \Delta\rho =  \sum_{\alpha} D_\alpha \rho D_\alpha^{\dagger}.
\end{equation}
For book-keeping, let us assume that one of the Kraus operators (denoting it as $D_0$) is close to the identity operator~$\mathbb{1}$. The other terms in Eq.~(\ref{Eq.koss1}) are of order $O(\Delta t)$. Up~to this order, we obtain
\begin{eqnarray}
   \label{Eq.koss2}
D_0 &=& \mathbb{1} + (L_{0} - \imath H) \Delta t, \nonumber \\
D_{\alpha} &=& L_{\alpha}\sqrt{ \Delta t},
\end{eqnarray}
where $\alpha \neq 0$. $L_{0}$ and $H$ are Hermitian. Expanding the terms in Eq.~(\ref{Eq.koss1}) gives
\begin{eqnarray}
   \label{Eq.koss3}
D_0 \rho D_0^{\dagger} &=& \rho + (L_{0}\rho +  \rho L_{0} - \imath H\rho +\imath \rho H)\Delta t + O(\Delta t^2),\nonumber \\
D_{\alpha} \rho D_{\alpha}^{\dagger} &=& L_{\alpha} \rho L_{\alpha}^{\dagger}  \Delta t.
\end{eqnarray}
Keeping terms to first order in $\Delta t$, we get
\begin{equation*}
\Delta\rho = \Bigg( \{L_{0}, \rho \} - \imath [H, \rho] + \sum_{\alpha} L_\alpha \rho L_\alpha^{\dagger}\Bigg) \Delta t. 
\end{equation*}
Hence
\begin{equation}
   \label{Eq.koss4}
\frac{d\rho}{dt} =  \{L_{0}, \rho \} - \imath [H, \rho] + \sum_{\alpha} L_\alpha \rho L_\alpha^{\dagger}.
\end{equation}
Note that $\{\cdot, \cdot\}$ represents the anti-commutator. Comparing with closed-system evolution, the Hamiltonian part of the open evolution must be $\tilde{H} = \hbar H$. Since the map is trace preserving, the trace of $\displaystyle \frac{d\rho}{dt} $ must be zero which gives us 
\begin{equation}
L_{0} = -\frac{1}{2} \sum_{\alpha}L_\alpha^{\dagger} L_{\alpha}.
\end{equation}
This gives us the Lindblad--Kossakowski--Gorini--Sudarshan master equation. 
\begin{equation}
   \label{Eq.koss5}
\frac{d\rho}{dt} =  - \imath \hbar [\tilde{H}, \rho] + \sum_{\alpha}\Bigg( L_\alpha \rho L_\alpha^{\dagger} - \frac{1}{2} \{L_\alpha^{\dagger} L_{\alpha}, \rho \}\Bigg).
\end{equation}
The $L_k$'s are usually called the Lindblad operators.

\section{Conclusions}

In this article, we have explained the different representations of finite-time maps describing open quantum evolution. It is shown how to switch between the various representations and the merits of the various ones are outlined. Discussion on the cases of several standard maps acting on a qubit, with the geometrical picture of mapping is also done. We also briefly discussed maps which are not completely positive, along with the differential version of the map.

\section*{Dedication}

We dedicate the present article to the memory of \mbox{E.~C.~G.~Sudarshan} who pioneered the concept of dynamical maps in open quantum evolution besides making many other profound contributions to theoretical physics.

\section*{Acknowledgment}

We thank Arul Lakshminarayan, Camille Lombard Latune and Vidya Kaipanchery for carefully reading the manuscript and for critical suggestions to improve the pedagogy. We also thank the editors for suggestions to improve the clarity of the presentation. This work is based upon research supported by the South African Research Chair Initiative (SARChI) of the Department of Science and Technology (DST) and the National Research Foundation (NRF).


\begin{thebibliography}{0}%
\makeatletter
\providecommand \@ifxundefined [1]{%
 \@ifx{#1\undefined}
}%
\providecommand \@ifnum [1]{%
 \ifnum #1\expandafter \@firstoftwo
 \else \expandafter \@secondoftwo
 \fi
}%
\providecommand \@ifx [1]{%
 \ifx #1\expandafter \@firstoftwo
 \else \expandafter \@secondoftwo
 \fi
}%
\providecommand \natexlab [1]{#1}%
\providecommand \enquote  [1]{``#1''}%
\providecommand \bibnamefont  [1]{#1}%
\providecommand \bibfnamefont [1]{#1}%
\providecommand \citenamefont [1]{#1}%
\providecommand \href@noop [0]{\@secondoftwo}%
\providecommand \href [0]{\begingroup \@sanitize@url \@href}%
\providecommand \@href[1]{\@@startlink{#1}\@@href}%
\providecommand \@@href[1]{\endgroup#1\@@endlink}%
\providecommand \@sanitize@url [0]{\catcode `\\12\catcode `\$12\catcode
  `\&12\catcode `\#12\catcode `\^12\catcode `\_12\catcode `\%12\relax}%
\providecommand \@@startlink[1]{}%
\providecommand \@@endlink[0]{}%
\providecommand \url  [0]{\begingroup\@sanitize@url \@url }%
\providecommand \@url [1]{\endgroup\@href {#1}{\urlprefix }}%
\providecommand \urlprefix  [0]{URL }%
\providecommand \Eprint [0]{\href }%
\providecommand \doibase [0]{http://dx.doi.org/}%
\providecommand \selectlanguage [0]{\@gobble}%
\providecommand \bibinfo  [0]{\@secondoftwo}%
\providecommand \bibfield  [0]{\@secondoftwo}%
\providecommand \translation [1]{[#1]}%
\providecommand \BibitemOpen [0]{}%
\providecommand \bibitemStop [0]{}%
\providecommand \bibitemNoStop [0]{.\EOS\space}%
\providecommand \EOS [0]{\spacefactor3000\relax}%
\providecommand \BibitemShut  [1]{\csname bibitem#1\endcsname}%
\let\auto@bib@innerbib\@empty
\end{thebibliography}%


\begin{thebibliography}{10}
\bibitem{breuer2007theory}
Breuer H-P, Petruccione F.
\emph{The Theory of Open Quantum Systems}. Oxford: Oxford University Press, 2007.

\bibitem{sudarshan61a}
Sudarshan ECG, Mathews PM, Rau J.
Stochastic dynamics of quantum-mechanical systems. \emph{Physical Review} 1961; \textbf{121}(3): 920--924. \href{http://dx.doi.org/10.1103/PhysRev.121.920}{\path{doi:10.1103/PhysRev.121.920}} 

\bibitem{kraus_general_1971}
Kraus K.
General state changes in quantum theory. \emph{Annals of Physics} 1971; \textbf{64}(2): 311--335. \href{http://dx.doi.org/10.1016/0003-4916(71)90108-4}{\path{doi:10.1016/0003-4916(71)90108-4}} 

\bibitem{dariano_bell_2000}
D'Ariano GM, Lo Presti P, Sacchi MF.
Bell measurements and observables. \emph{Physics Letters A} 2000; \textbf{272}(1): 32--38. \href{http://arxiv.org/abs/quant-ph/0005121}{\path{arXiv:quant-ph/0005121}}, \href{http://dx.doi.org/10.1016/S0375-9601(00)00410-2}{\path{doi:10.1016/S0375-9601(00)00410-2}} 

\bibitem{stinespring55}
Stinespring WF.
Positive functions on C*-algebras. \emph{Proceedings of the American Mathematical Society} 1955; \textbf{6}(2): 211--216. \href{http://dx.doi.org/10.2307/2032342}{\path{doi:10.2307/2032342}} 

\bibitem{choi1975}
Choi M-D.
Completely positive linear maps on complex matrices. \emph{Linear Algebra and its Applications} 1975; \textbf{10}(3): 285--290. \href{http://dx.doi.org/10.1016/0024-3795(75)90075-0}{\path{doi:10.1016/0024-3795(75)90075-0}} 

\bibitem{chuang_prescription_1997}
Chuang IL, Nielsen MA.
Prescription for experimental determination of the dynamics of a quantum black box. \emph{Journal of Modern Optics} 1997; \textbf{44}(11--12): 2455--2467. \href{http://arxiv.org/abs/quant-ph/9610001}{\path{arXiv:quant-ph/9610001}}, \href{http://dx.doi.org/10.1080/09500349708231894}{\path{doi:10.1080/09500349708231894}} 

\bibitem{beth_ruskai_analysis_2002}
Ruskai MB, Szarek S, Werner E.
An analysis of completely-positive trace-preserving maps on $\mathcal{M}_2$. \emph{Linear Algebra and its Applications} 2002; \textbf{347}(1): 159--187. \href{http://dx.doi.org/10.1016/S0024-3795(01)00547-X}{\path{doi:10.1016/S0024-3795(01)00547-X}} 

\bibitem{jordan_dynamical_1961}
Jordan TF, Sudarshan ECG.
Dynamical mappings of density operators in quantum mechanics. \emph{Journal of Mathematical Physics} 1961; \textbf{2}(6): 772--775. \href{http://dx.doi.org/10.1063/1.1724221}{\path{doi:10.1063/1.1724221}} \url{https://doi.org/10.1063/1.1724221}

\bibitem{jordan_dynamical_1962}
Jordan TF, Pinsky MA, Sudarshan ECG.
Dynamical mappings of density operators in quantum mechanics. II. Time dependent mappings. \emph{Journal of Mathematical Physics} 1962; \textbf{3}(5): 848--852. \href{http://dx.doi.org/10.1063/1.1724298}{\path{doi:10.1063/1.1724298}} \url{https://doi.org/10.1063/1.1724298}

\bibitem{simmons_completely_1981}
Simmons RF, Park JL.
On completely positive maps in generalized quantum dynamics. \emph{Foundations of Physics} 1981; \textbf{11}(1): 47--55. \href{http://dx.doi.org/10.1007/bf00715195}{\path{doi:10.1007/bf00715195}} 

\bibitem{raggio_remarks_1982}
Raggio GA, Primas H.
Remarks on ``On completely positive maps in generalized quantum dynamics''. \emph{Foundations of Physics} 1982; \textbf{12}(4): 433--435. \href{http://dx.doi.org/10.1007/bf00726787}{\path{doi:10.1007/bf00726787}} 

\bibitem{simmons_another_1982}
Simmons RF, Park JL.
Another look at complete positivity in generalized quantum dynamics: Reply to Raggio and Primas. \emph{Foundations of Physics} 1982; \textbf{12}(4): 437--439. \href{http://dx.doi.org/10.1007/bf00726788}{\path{doi:10.1007/bf00726788}} 

\bibitem{pechukas_reduced_1994}
Pechukas P.
Reduced dynamics need not be completely positive. \emph{Physical Review Letters} 1994; \textbf{73}(8): 1060--1062. \href{http://dx.doi.org/10.1103/PhysRevLett.73.1060}{\path{doi:10.1103/PhysRevLett.73.1060}} 

\bibitem{alicki_comment_1995}
Alicki R.
Comment on ``Reduced dynamics need not be completely positive''. \emph{Physical Review Letters} 1995; \textbf{75}(16): 3020. \href{http://dx.doi.org/10.1103/PhysRevLett.75.3020}{\path{doi:10.1103/PhysRevLett.75.3020}} 

\bibitem{pechukas_pechukas_1995}
Pechukas P.
Pechukas replies. \emph{Physical Review Letters} 1995; \textbf{75}(16): 3021. \href{http://dx.doi.org/10.1103/PhysRevLett.75.3021}{\path{doi:10.1103/PhysRevLett.75.3021}} 

\bibitem{jordan_dynamics_2004}
Jordan TF, Shaji A, Sudarshan ECG.
Dynamics of initially entangled open quantum systems. \emph{Physical Review A} 2004; \textbf{70}(5): 052110. \href{http://arxiv.org/abs/quant-ph/0407083}{\path{arXiv:quant-ph/0407083}}, \href{http://dx.doi.org/10.1103/PhysRevA.70.052110}{\path{doi:10.1103/PhysRevA.70.052110}} 

\bibitem{shabani_maps_2009}
Shabani A, Lidar DA.
Maps for general open quantum systems and a theory of linear quantum error correction. \emph{Physical Review A} 2009; \textbf{80}(1): 012309. \href{http://arxiv.org/abs/0902.2478}{\path{arXiv:0902.2478}}, \href{http://dx.doi.org/10.1103/PhysRevA.80.012309}{\path{doi:10.1103/PhysRevA.80.012309}} 

\bibitem{buscemi_complete_2014}
Buscemi F.
Complete positivity, Markovianity, and the quantum data-processing inequality, in the presence of initial system-environment correlations. \emph{Physical Review Letters} 2014; \textbf{113}(14): 140502. \href{http://arxiv.org/abs/1307.0363}{\path{arXiv:1307.0363}}, \href{http://dx.doi.org/10.1103/PhysRevLett.113.140502}{\path{doi:10.1103/PhysRevLett.113.140502}} 

\bibitem{shaji_whos_2005}
Shaji A, Sudarshan ECG.
Who's afraid of not completely positive maps? \emph{Physics Letters A} 2005; \textbf{341}(1): 48--54. \href{http://dx.doi.org/10.1016/j.physleta.2005.04.029}{\path{doi:10.1016/j.physleta.2005.04.029}} 

\bibitem{carteret_dynamics_2008}
Carteret HA, Terno DR, {\.Z}yczkowski K.
Dynamics beyond completely positive maps: some properties and applications. \emph{Physical Review A} 2008; \textbf{77}(4): 042113. \href{http://dx.doi.org/10.1103/PhysRevA.77.042113}{\path{doi:10.1103/PhysRevA.77.042113}} 

\bibitem{Chruscinski-2014}
Chru{\'s}ci{\'n}ski D, Sarbicki G.
Entanglement witnesses: construction, analysis and classification. \emph{Journal of Physics A: Mathematical and Theoretical} 2014; \textbf{47}(48): 483001. \href{http://arxiv.org/abs/1402.2413}{\path{arXiv:1402.2413}}, \href{http://dx.doi.org/10.1088/1751-8113/47/48/483001}{\path{doi:10.1088/1751-8113/47/48/483001}} 

\bibitem{rivasprl}
Rivas {\'A}, Huelga SF, Plenio MB.
Entanglement and non-Markovianity of quantum evolutions. \emph{Physical Review Letters} 2010; \textbf{105}(5): 050403. \href{http://arxiv.org/abs/0911.4270}{\path{arXiv:0911.4270}}, \href{http://dx.doi.org/10.1103/PhysRevLett.105.050403}{\path{doi:10.1103/PhysRevLett.105.050403}} 

\bibitem{rivasreview}
Rivas {\'A}, Huelga SF, Plenio MB.
Quantum non-Markovianity: characterization, quantification and detection. \emph{Reports on Progress in Physics} 2014; \textbf{77}(9): 094001. \href{http://arxiv.org/abs/1405.0303}{\path{arXiv:1405.0303}}, \href{http://dx.doi.org/10.1088/0034-4885/77/9/094001}{\path{doi:10.1088/0034-4885/77/9/094001}} 

\bibitem{gorini_completely_1976}
Gorini V, Kossakowski A, Sudarshan ECG.
Completely positive dynamical semigroups of N‐level systems. \emph{Journal of Mathematical Physics} 1976; \textbf{17}(5): 821--825. \href{http://dx.doi.org/10.1063/1.522979}{\path{doi:10.1063/1.522979}} 

\bibitem{lindblad76}
Lindblad G.
On the generators of quantum dynamical semigroups. \emph{Communications in Mathematical Physics} 1976; \textbf{48}(2): 119--130. \href{http://projecteuclid.org/euclid.1103899849}{\path{Euclid:1103899849}}, \href{http://dx.doi.org/10.1007/bf01608499}{\path{doi:10.1007/bf01608499}} 

\end{thebibliography}
\end{document}